\documentclass[reprint,amsmath,amssymb]{revtex4-2}

\usepackage{graphicx}
\usepackage{dcolumn}
\usepackage{bm}
\usepackage{hyperref}

\begin{document}
\author{Diego J. Monterrubio-Chanca}
\author{Guido Falk von Rudorff}
\affiliation{Institut f\"ur Chemie, Universit\"at Kassel, 34109 Kassel}
\affiliation{Center for Interdisciplinary Nanostructure Science and Technology (CINSaT), Heinrich-Plett-Straße 40, 34132 Kassel}

\title{Representative Random Sampling of Chemical Space}

\date{\today}
\begin{abstract}

The overwhelming majority of molecules remains unexplored. This is mostly due to the sheer number of them, which prohibits any enumeration of chemical space, the set of all such molecules. In practice, only subsets of chemical space are considered, but those subsets exhibit substantial bias, prohibiting data-driven characterization of chemical space itself. In this work, we provide a method produce unbiased representative random samples of chemical space without enumeration of constituent molecules and to estimate the number of molecules in any custom chemical space. The approach is applicable to molecules which can be represented as graph and runs efficiently even for molecules of 30 atoms. We use it to estimate the representativeness of current databases with respect to their underlying chemical space and to establish a necessary criterion for a lower bound of database sizes to be representative of an underlying chemical space.

\end{abstract}
\maketitle

\maketitle
\begin{figure*}[htbp]
    \centering
 \includegraphics[width=\textwidth,
  trim=6cm 4cm 2.5cm 4cm, clip]{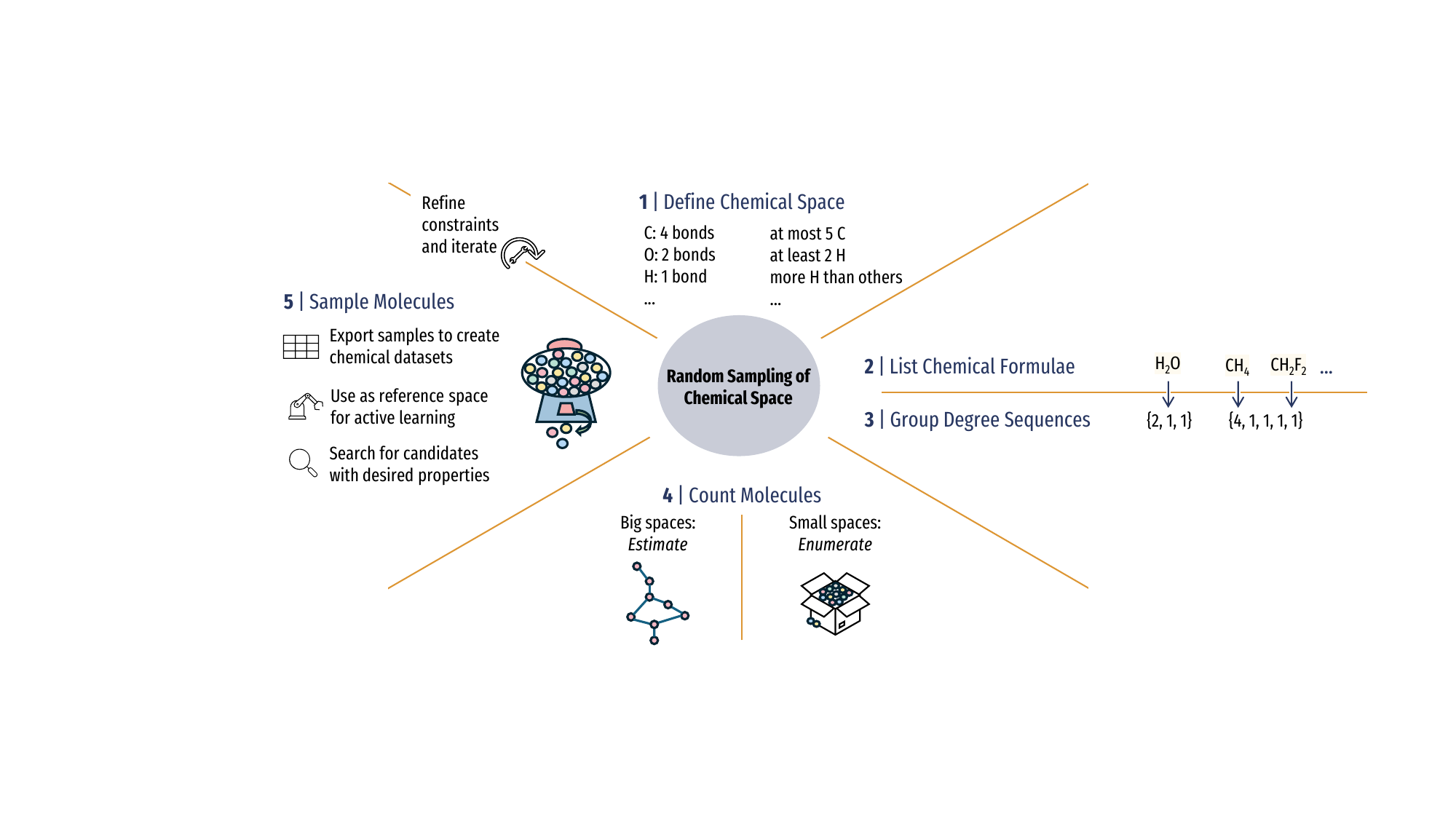}
    \caption{Overview of the random sampling procedure and applications. From a well defined chemical space based on valency and stoichiometry restrictions (1), all chemical formulae are obtained by integer partition (2) and grouped by degree sequence (3). For each of those, the total size is estimated (4) which allows for appropriate weighted random sampling(5). }
    \label{fig:flowchart}
\end{figure*}

\section{Introduction}

Chemical space is the set of all stable compounds from which we can choose molecules of favorable properties and therefore the search space for compound design\cite{Reymond2015,Montavon2013,keith2021combining,Medina-Franco2022}. For practical reasons, we typically consider only a subset of that space, even when statements about general structure-property relationships are made. If this subset is not representative of chemical space, it introduces a bias which propagates to the conclusions drawn from the subset analysis. Although quantum chemistry datasets\cite{Hoja2021,Ganscha2025,Smith2017a,Smith2020,Blum2009,Ruddigkeit2012,Ramakrishnan2014,Zheng2009,Balcells2020,Nakata2020,Schreiner2022,Zhao2023,Isert2022,Ehlert2025,Chakraborty2025} grow larger and larger\cite{Ullah2024,Levine2025}, any self-similarity of the data points they cover does not necessarily reduce the bias of the data distribution\cite{Hossain2025,LopezPerez2025}. These self-similarities are also introduced by the whole tool chain that is used in generating the databases\cite{Duan2019,Heinen_2020}. Therefore, it is highly desirable to assess chemical space in an provably unbiased manner

Unbiased exploration would improve our intuition on chemical processes and modelling. Textbook concepts such as electron donating/withdrawing groups, the Hammond postulate\cite{Hammond1955} or Hückel's rule\cite{Hueckel1931} are derived from experimental data (which by the nature of the intricate experimental process is self-similar) or theoretical models. As such this might constitute overfitting towards the tiny section of chemical space we have observed so far and may break down systematically for other regions\cite{Heinen2021, Arteca1988,Donahue2001,VanNyvel2025}, failing to generalize to an appreciable number of systems\cite{Apriliyanto2021}.

Considering main goals of modern machine learning methods, that is efficiency, accuracy, scalability, and transferability (EAST)\cite{Huang2023}, unbiased reference data connects efficiency and transferability. The smaller the required amount of training data, i.e. the higher the data efficiency, and the more evenly balanced this data is over the target domain, and the higher the transferability which can be reasonably expected.

Exploring the vastness of chemical space is relevant for a wide range of applications, such as drug design\cite{Lavecchia2024,Sadybekov2023} or catalysts\cite{Griego2021,Tan2025,Mace2024}, novel materials \cite{GomezBombarelli2018, Cao2025} and molecular properties prediction \cite{Liu2023, Yang2025}. Given the extent of exponentially increasing chemical space, its vast majority is unexplored and probably will remain unexplored, which makes it even more important to traverse the options most efficiently. Researchers have addressed this substantial challenge by forward design based on predicting individual molecular properties with a wide range of methods\cite{Sun2025a,Suman2025,Smith2017b,Stuke2019,Serrano-Morras2025} and inverse design using generative models\cite{Vogt2023,Anstine2023}. In forward predictions, a core challenge is to reliably assess the generalization error, which is typically done using hold out data sets, assuming that the underlying database is representative of the problem space, since otherwise aggregate measures do not apply to true out-of-sample data. For generative models, an open question is whether their image space captures the full target domain, or whether they are most likely to reproduce compounds similar to the training data or how to restrict to meaningful output spaces\cite{Langevin2023,Sosnin2025}. While large training data sets were a necessary requirement for machine learning of molecular properties, most early data sets make assumptions simplifying their generation and include artificial biases towards compounds which have better support from cheminformatics toolchains\cite{Ramakrishnan2014} or quantum chemistry software\cite{Duan2019a}. Those restrictions however skip interesting systems which would commonly be assumed to be unstable, such as cubane \cite{Bugg1964}, hexamethylbenzene dication \cite{Malischewski2017}, argon fluorohydride \cite{Khriachtchev2000}, hexanitrogen \cite{Qian2025}, C$_{48}$  (cyclo[48]carbon) \cite{Gao2025}. This impacts or hinders the discovery of of the so-called unknown unknowns, i.e. new effects or physical phenomena.

One approach of systematic exploration of chemical space is based on full enumeration. Over decades, many tools have been developed such as MOLGEN\cite{Benecke1997}, OMG\cite{Peironcely2012}, PMG\cite{Jaghoori2013}, surge\cite{McKay2022}, Maygen\cite{Yirik2021}, and enu\cite{Rieder2023}. What they have in common is that they systematically enumerate the space of all feasible molecules by constraining themselves to those bond topologies which are in line with standard valence bond rules and then systematically generate molecular graphs. The major achievement of those codes is that duplicate checks are not required as the generation itself already guarantees uniqueness. However, even those approaches are limited in molecular size since chemical space is scaling even further than the few ten atoms that can commonly be treated. To combat the exponential scaling of chemical space, most of these tools allow the definition of further restriction such as avoiding certain features of molecular graphs or further constraining bond topology or subgroups. This has been very helpful in practical application, but only marginally increases the size of the molecules that can be enumerated systematically while introducing the aforementioned biases. In the context of structure elucidation, stochastic approaches have been pursued\cite{Faulon1994} with a focus on identifying possible target compounds if molecular fragments are known, which does not allow for random sampling of a chemical space. More recently, larger chemical diversity became the goal of molecular generators\cite{Gasevic2025}.

Moreover, these methods cannot be parallelized easily, even though some have been developed with parallelization in mind\cite{Jaghoori2013,McKay2022}. Since in the systematic enumeration the problem cannot be subdivided evenly, even approximately random uniform sampling is infeasible with those methods, unless generating a complete list of all molecules would be feasible.

The infeasibility of full enumeration renders it necessary to employ probabilistic approaches for unbiased predictions over a chemical space constrained by domain-specific physical limits such as element variety, molecular size and stoichiometries instead of software support or practical convenience. For example, local refinement in chemical space can be carried out reliably using Monte Carlo methods\cite{Karandashev2023a}. However global exploration poses a considerable challenge: accessing all valid molecules requires long, sequential transitions, which are computationally expensive and difficult to parallelise. Ensuring detailed balance and unbiased sampling further complicates the process, leaving full exploration through classical Markov Chain Monte Carlo (MCMC) remains impractical. An unbiased random sampler, however, could parallelize MCMC by providing a more diverse set of starting points. 

In this work, we present a method that generates approximately uniform random samples spanning a chemical space. We only require that the chemical space admits a representation of a molecule as a finite molecular graph, which excludes essentially the same application domains where vanilla SMILES representations are inapplicable, such as polymers\cite{Krenn2022}. This approach enables scalable and representative exploration of chemical space, suitable for applications in design, learning, and structural analysis.

\section{Methods}
Our method \textit{Representative Random Sampling}  (RRS) produces a sequence of molecules which are approximately uniformly randomly selected from a given chemical space spanned by all molecular graphs which satisfy valence bond rules. While this is not applicable to all chemical spaces it is commonly used\cite{Krenn2022}.

In RRS, we consider atoms of different valences as different \textit{atom types}, e.g. phosphorus of valence 3 and phosphorus of valence 5 would be different atom types. The atom types form the set $A$ of pairs $(e,v)$ with element $e$ and valence $v$. All atoms of same valence are of the same \textit{valence type}, e.g. hydrogen and all halogens. In the following, $V \equiv \{a_2|a\in A\}$ is the ordered set of valence types.
For each valence type $v$, there are potentially multiple atom types $A(v) \equiv \{a \in A| a_2 = v\}$, counted by the \textit{valence type multiplicity} $|A(v)|$. This allows us to simplify the problem considerably, since the number of possible molecules that can be formed given a set of atoms only depends on their valence type not on their element. While \textit{enumerating} graphs via graph counting methods is implemented in many codes, it is not fast enough for the problem size at hand where computing or storing the list of molecular graphs is completely infeasible. We address this by first \textit{estimating} the total number of molecular graphs for each sum formula within a search space and then uniformly randomly sampling from that search space by selecting a chemical formula followed by an Markov Chain Monte Carlo sampler within that chemical formula. The key difficulty here is to reliably estimate how many molecular graphs exist for any given chemical formula.

\subsection{Obtaining all chemical formulae}
In principle, obtaining all chemical formulae within a chemical space is an integer partition  problem where a certain number of atoms is distributed over all possible elements allowed in the chemical space. Due to the sparsity structure of those partitions, a nested approach is substantially more efficient than direct partitioning. We first find the set of constitutions $C$ for molecules in a given chemical space by repeated integer partitions. A single constitution is defined as a multiset of atom types, i.e. element and valencies. 

Let $p(c, B)$ be the integer partition of the ordered set $B$ into $c$ parts $\{(x_1, \dots, x_{|B|}) | \sum_i x_i = c\}$. For each possible number of atoms $N_a$ in the chemical space, we obtain a set of partitions
\begin{align}
    O(N_a)\equiv \{q\in p(N_a, V) | M_V(q) \textrm{ even} \wedge S(q) \}
\end{align}
where $M_V(q)\equiv \sum_i q_iv_i$ is the sum of all degrees and $S(q)$ is true of all bonds can be saturated given the valencies\cite{McKay2022}. $M_V(q)$ needs to be even such that a graph is feasible. At this stage, only the valencies are taken into account but not the valence type (e.g. so far monovalent hydrogen and fluorine would not be distinguished). 

We now find all possible constitutions $C(N_a)$ by taking the Cartesian product $c$ of all partitions within each valence type, i.e. by finding all constitutions including the atom type information
\begin{align}
    C(N_a) &\equiv\bigcup_i \left\{c(x)|x\in O(N_a)\right\}\\
    c(x)&\equiv  \prod_i\{a^{y_i} | y_i\in p(x_i, A(v_i)) \wedge a\in A(v_i) \}
\end{align}

This two-stage approach is computationally much more efficient yet otherwise identical to directly partitioning $N_a$ atoms into the corresponding elements from the chemical space, since only unique cases are generated: e.g. CF$_4$ and CH$_4$ correspond to the same element in $O(5)$, but become different constitutions in $C(5)$. 

For each constitution, there are potentially many connected loop-free multigraphs where vertices are labeled with the chemical element. These graphs all have the same degree sequence $d$ where each vertex (an atom) is also assigned an element. To exploit symmetries (such as the CF$_4$ case above), we only consider unique \textit{protomolecules} where fictitious element labels are placeholders for all elements of compatible atom types, but including the distinction from the constitution. For example, using superscripts as valencies CH$_2$F$_2$ and CH$_2$Cl$_2$ both form protomolecules of family X$^4$Y$^1_2$Z$^1_2$, and therefore are distinct from CF$_4$ and CH$_4$ which belong to protomolecule family X$^4$Y$^1_4$.

\subsection{Estimating molecule counts}
For small number of atoms (typically up to about 10), direct systematic enumeration of all protomolecules is feasible. In this work, we use \textit{surge}\cite{McKay2022}; all resulting counts are distributed with our software packages nablachem. Beyond this realm, we obtain approximate counts using results from network topology. First, we introduce a new graph, the \textit{universe} $U(d)$, which has one vertex for each protomolecule that can be obtained from a labeled degree sequence $d$. In this universe, two vertices are connected if and only if their corresponding vertices have minimal graph edit distance. We then use the results for small world networks\cite{Watts1998} which states that the average length $l_G$ of the shortest path between two vertices can be related approximately to the total number of vertices in that graph:
\begin{align}
 l_G\sim \log |U(d)|   \label{eqn:smallworldnetwork}
\end{align}
This assumes that molecular universes can be considered a small world network which numerical results will justify. The main advantage of this approach is that we only need the average distance between elements of this universe, meaning we do not need to enumerate all of the molecules that have the same labeled degree sequence, but rather a tiny subset thereof, since the average of their pairwise distances converges quickly. This is the key step enabling the reliable estimation of molecular counts in this work.  
Obtaining $l_G$ by randomly sampling protomolecules and finding their edit distance then allows to estimate the total number of protomolecules $|U(d)|$ for that labeled degree sequence $d$. This is computationally feasible until about 20 atoms.

For any pure degree sequence, i.e. where the maximum valence multiplicity is 1, the number of distinct protomolecules is minimal amongst all degree sequences which have the same valencies but with larger multiplicities. E.g. if we have two atoms of same valency and identical element label (pure degree sequence), that yields one protomolecule. If we have two atoms of same valency and different element labels (non-pure degree sequence), we can now have three different protomolecules (with the element labels one-one, one-other and other-other). If we assume that for large and complex molecules which contain many branches and asymmetries, isomorphisms are rare, we expect that almost always each atom site is unique in chemical environment. Therefore, the number of possible protomolecules $N_P(d)$ upon introducing more elements with the same valency can be approximated by applying combinatorial counting to groups in the labeled degree sequence $d$:
\begin{align}
    N_P(d) = \prod_v \prod_i \binom{\sum_{j\ge i} c_j}{c_i}\label{eqn:pureisenough}
\end{align}
where $v$ runs over all valencies in the degree sequence, and $i$ and $j$ run over all elements of that given valency $v$. For each $i$-th element, $c_i$ is the number of atoms of that element. Note that for unlabeled degree sequences, $N_P(d)=1$. The asymptotic behavior of protomolecule counts\cite{Greenhill2013} suggests that the total number of protomolecules should be inversely related to the total number of bonds $M(d)$. This yields
\begin{align}
    l_G(d) = \left(1+\left[\sum_i d_i \right]^{-1}\log N_P(d)\right)l_G(d_U) 
\end{align}
as expression to estimate the average path length $l_G(d)$ for an arbitrary degree sequence if the average path length $l_G(d_U)$ for the corresponding pure degree sequence is known. The number of protomolecules is then obtained from $l_G$ as it is for smaller molecules. This is computationally feasible until about 30 atoms.

Finally, we have reached the realm of asymptotic scaling, where formal statements about the number of protomolecules exist\cite{Greenhill2013}. This allows to obtain the scaling behaviour of a quantity proportional to $l_G$ of any unlabeled degree sequence. Using the permutation logic of the previous paragraph allows to then obtain $l_G$ for any labeled degree sequence. This is extremely cheap and can be applied to arbitrarily large molecules. The bottleneck then becomes the enumeration of all sum formulas from the integer partitions in the first place.

\begin{table}[]
    \centering
    \begin{tabular}{c|c|c|c}
        Space & Valence & Multiplicity & Example\\\hline\hline
        A&1 & 5 & F, H, Cl, Br, I\\
        & 2 & 2 & O, S\\
        & 3 & 2 & N, P\\
        & 4 & 3 & C, S, Si\\
        & 5 & 2 & N, P\\
        & 6 & 1 & S\\\hline
        B&1 & 5 & F , H, Cl, Br, I\\
        & 2 & 2 & O\\
        & 3 & 2 & N, P\\
        & 4 & 1 & C\\
        & 5 & 1 & P
    \end{tabular}
    \caption{Chemical spaces used in this work. Element labels are illustrative only, since any labeling of equivalently bonding elements yields the same protomolecule set.}
    \label{tab:spaces}
\end{table}

\begin{table}[]
    \centering
    \begin{tabular}{c|c|c|c}
        Space & Exact & Base & Pure \\\hline\hline
        A & 3-10 atoms & 10-15 atoms & 10-20 atoms\\
        B & 3-10 atoms & 10-22 atoms & 10-31 atoms
    \end{tabular}
    \caption{System size coverage of the data in this work for each chemical space. For the domain labeled \textit{exact}, the correct weights are known from enumeration, so the sampling is guaranteed to be uniform. For \textit{base} data, the average path length has been found heuristically from explicit sampling of all labeled degree sequences. For \textit{pure} domains, this has been done for all unlabeled degree sequences only. Above the specified domains, asymptotic scaling relations are employed.  \textit{base} is a strict superset of \textit{pure} data.}
    \label{tab:spacedata}
\end{table}

\subsection{Estimating the Minimal Graph Edit Distance}
The average path length of the shortest path between two vertices in the universe is  the graph edit distance, meaning the minimal sets of edits that have to be done to one molecular graph in order to obtain the other. While the graph edit distance cannot be determined in polynomial time, efficient heuristics exist. In this work, we measure the graph edit distance as the minimal Wasserstein metric between the all permutations of the adjacency matrices of two molecules. For small molecules with fewer than 100 permutations, all permutations are considered, so the minimisation is exact. For larger ones, we use the minimal Wasserstein metric obtained from five different heuristics: i) random shuffling, ii) repeatedly search for a pairwise switch of atom indices which minimizes the metric from a random initial value, iii) treating the permuation as quadratic assignment problem (QAP) with the 2-opt\cite{Croes1958} algorithm allowing all permutations, iv) as repeated QAP with permutations freezing all but one chemical element sequentially in one scan or v) as QAP with permutations freezing all but one chemical element in multiple scans and vi) employing a depth-first graph edit distance optimizer\cite{Abu-Aisheh2015}. Each such heuristic has been called about 50-100 times depending on graph size.

We empirically find that for different parts of chemical space these heuristics have different efficiency but did not analyze this further as current computational cost is acceptable. 

\begin{figure*}
    \centering
    \includegraphics[width=\textwidth]{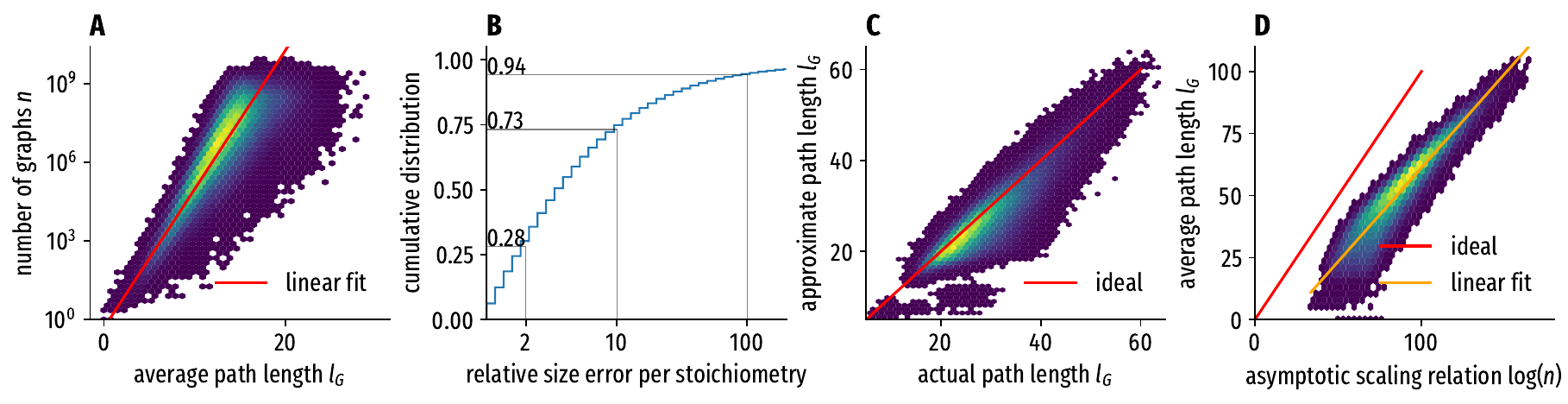}

    \caption{Accuracy of the correlations exploited in this work. 
    \textbf{A}: Obtaining counts $n$ of protomolecules from the average path length $l_G$ in the universe graph for all 285,656 stoichiometries with exactly 26,076,359,902,577 molecular graphs in our database. Histogram of the correlation between $l_G$ and $n$ together with the expected linear fit. 
    \textbf{B}: Relative error of the estimated and the actual number of graphs for a given stoichiometry as cumulative histogram: correct order of magnitude is reached for 73\% of the cases.
    \textbf{C}: Estimated average path length vs actual average path length for all 553,132 non-pure labeled degree sequences in our database. Note that there is no fitting involved in this step.
    \textbf{D}: Accuracy of estimating the logarithm of the number of molecules with pure degree sequences from calibrating the asymptotic scaling relations in Eqn.~\ref{eqn:asymptotic}. The density histogram covers all 148,620 pure degree sequences with more than 20 atoms in our database.}
    \label{fig:exact-base}
\end{figure*}

\section{Results}
\subsection{Estimating molecule counts}
The core idea to estimate the size of chemical spaces from the average path length following the small world network results in eqn.~\ref{eqn:smallworldnetwork} first needs validation. To this end, we \textit{enumerated} small molecular graphs using \textit{surge}\cite{McKay2022}. Table~\ref{tab:spaces} shows the chemical spaces considered and Table~\ref{tab:spacedata} shows the sizes of the molecules for which there is pre-computed data available. It is important to note that the element labels are arbitrary in this work: one can easily replace one element forming a single bond with another element forming a single bond without any additional computation. The pre-computed databases are limited by how many \textit{distinct} elements can be used of a given valence, the \textit{valence multiplicity}.

The \texttt{nablachem.space}\cite{Banjafar2025c} python package contains all this pre-computed data and automatically chooses the most accurate method available when queried. It also contains the code to extend the databases in this work to arbitrary chemical spaces.

Fig.~\ref{fig:exact-base}A shows the correlation between the number of molecules $n$ for a given chemical formula and the average path length $l_G$ as determined by our heuristic. We find that there is a strong correlation between $\log(n)$ and $l_G$, indicating that eqn.~\ref{eqn:smallworldnetwork} is applicable to the data at hand. This is a key results which allows to estimate the number of molecules for a given chemical formula without any enumeration by parametrizing the proportionality from eqn.~\ref{eqn:smallworldnetwork}. When fitting to the exact counts for all 285,656 unique degree sequences covering 26,076,359,902,577 molecular graphs in our database in Fig.~\ref{fig:exact-base}A, we obtain the following approximation
\begin{align}
    \log |U(d)| \approx 1.220 l_G -0.7295 \label{eqn:universefit}
\end{align}
We find that extending the sampling steps of our heuristic does not reduce the average path length significantly. We therefore consider the deviations in the histogram to be the consequence of the extent to which the small world network approximation is valid. Since the fit is performed on logarithmized data, it is well-balanced across the widely different number of molecular graphs in each stoichiometry: Fig.~\ref{fig:exact-base}B shows the histogram of the relative errors made in the size estimation between the real enumerated number of graphs and the estimate from eqn.~\ref{eqn:universefit}. Our method estimates the correct order of magnitude for each stoichiometry in 73\,\% of the cases. Since some stoichiometries are underestimated (the region above the linear fit in Fig.~\ref{fig:exact-base}) and some are overestimated (the region below the linear fit), the random sampling from cumulative distributions remains balanced throughout chemical space. The strong limitation arising for our method however is that there is a systematic error for any \textit{single} stoichiometry which does not average out.

Since at some point even enumerating all degree sequences of interest becomes prohibitively expensive, eqn.~\ref{eqn:pureisenough} was the consequence of a simple combinatorial argument for estimating the number of unique graphs arising from non-pure degree sequences if only the size of the pure degree sequence (or, equivalently, its average path length) is known. This hypothesis is probable, since a random molecular graph is unlikely to be symmetric, and could be confirmed in the data: Fig.~\ref{fig:exact-base}C shows the correlation between the average path length using the explicit minimum edit graph heuristic above and the results of applying eqn.~\ref{eqn:pureisenough}. Note that there is no fitting involved in this step. This greatly reduces the computational complexity from estimating the size of all degree sequences to the one of estimating the size of pure degree sequences of which there are much fewer.

Going towards even larger molecules, the heuristic determination of the average path length becomes prohibitively expensive. Fortunately, asymptotic scaling relations for the number of graphs in this case are available\cite{Greenhill2013} (Theorem 1.1) and are reproduced here for simplicity:
\begin{align}
    G = &\frac{M!}{(M/2)!2^{M/2}k_1!\cdots k_n!}\nonumber\\
    &\exp\Bigg(\left(y_1-\frac{1}{2}\right)\frac{M_2}{M}+\left(x_2-\frac{1}{2}\right)\frac{M_2^2}{2M^2} + \frac{M_2^4}{4M^5}\nonumber\\
    &-\frac{M_2^2M_3}{2M^4} + \left(x_3-x_2+\frac{1}{3}\right)\frac{M_3^2}{2M^3} + \mathcal{O}(k_\text{max}^3/M)\Bigg)\label{eqn:asymptotic}\\
    M_r  =& \sum_i^n [k_i]_r
\end{align}
where $n$ is the length of the degree sequence $\mathbf{k}$, $[\cdot]_r$ denotes the falling factorial and $x_i=1$ if $i$ is allowed as non-loop edge multiplicity (otherwise 0) and $y_i=1$ if $i$ is allowed as loop-edge multiplicity (otherwise 0). Here, this means all $x_i=1$ and all $y_i = 0$ for $i > 0$. 

Since this is an asymptotic scaling relation, the prefactor is unknown. In this work, we calibrate the scaling relation to actual data by a linear fit of $l_G$ to $\log(G)$ on all 148,620 pure degree sequences with more than 20 atoms for which $l_G$ is known (see Fig.~\ref{fig:exact-base}) and obtain
\begin{align}
    l_G \approx 0.7561 \log(G) -14.40
\end{align}
this simple model performs remarkably well and remains feasible as long as listing the integer partitions of the total number of atoms is affordable. The non-pure degree sequence count then is estimated using eqn.~\ref{eqn:pureisenough} as before.

When adding $t$ Hydrogens (or other monovalent elements) to a fixed stoichiometry, eventually the number of molecular graphs has to decrease, as the double bond equivalents of the graphs approach zero. The asymptotic scaling relation does not take this effect into account: it would grow monotonously instead, even once graphs become infeasible due to the negative double bond equivalent. Comparing to the cases for which we know the exact graph count from our database, we find that the relation tends to overestimate by $t!$, so we correct by this amount. Naturally, the extreme end of the molecule count estimates are impossible to calibrate, so we recommend to limit the use of the asymptotic scaling relation once it yields a large variance for comparable stoichiometries.

\subsection{Sampling molecular graphs}
Once for each chemical formula the total number of molecular graphs is estimated, we can finally sample molecules for a particular chemical formula which has been selected with a probability proportional to its size in the selected chemical space. We implement this following a Markov-Chain-Monte Carlo (MCMC) approach which allows to obtain a random multigraph of a given degree sequence\cite{Greenhill2015,jamesross22023}. Since the space of molecules to sample from is restricted to those of the same degree sequence, sampling the distribution converges quickly even for large molecular graphs.

In practice, there are two main limitations: i) MCMC will find each member of the automorphism group $\textrm{Aut}(G)$ with equal probability not only each unique graph $G$ and ii) random multigraphs are not necessarily connected. We deal with the former issue via rejection sampling by accepting a potential MCMC result with probability $1/|\textrm{Aut}(G)|$ which is an acceptable inefficiency, since it only becomes costly if a degree sequence would ever be dominated by symmetric graphs. The latter problem is much more severe in practice: if for a given degree sequence the total number of connected graphs is too small, the likelihood of a random graph being connected drops rapidly. While our current implementation keeps trying and just samples this region of chemical space slowly, other random graph generation methods could potentially used to improve sampling speed of those domains. Since this does not limit sampling accuracy in all other regions of chemical space, we consider numerical efficiency of this issue a topic for future work.

The sampling code is available in \texttt{nablachem.space}\cite{Banjafar2025c} as well can be run interactively on \texttt{random-molecule.org}.

\begin{figure*}
    \centering
    \includegraphics[width=\textwidth]{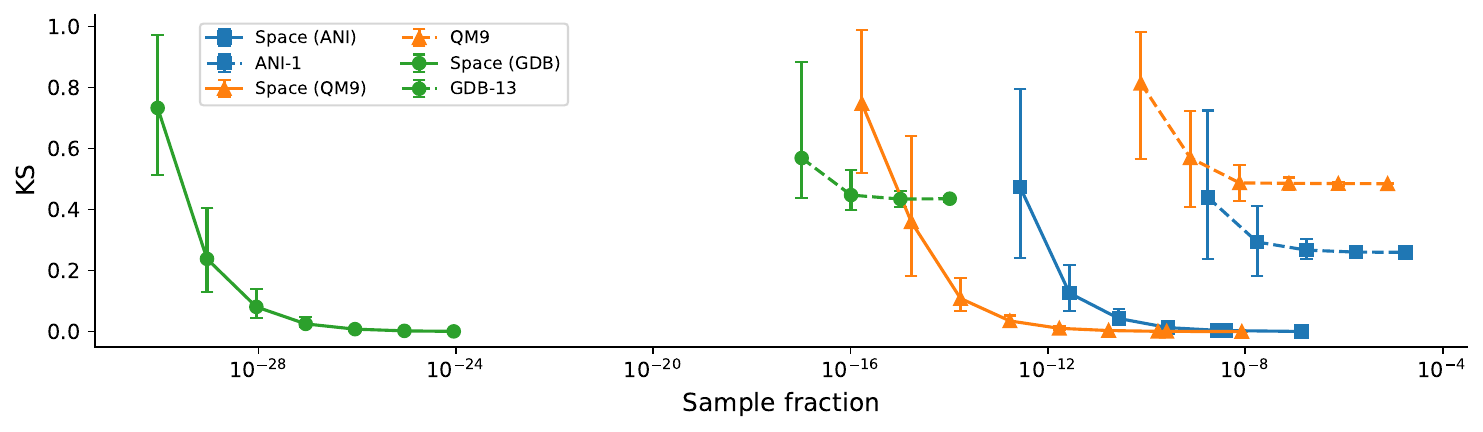}
    \caption{Kolmogorov–Smirnov (KS) statistic as a function of the sampled fraction of the corresponding chemical space. Solid lines represent comparisons between the full space and randomized subspaces thereof. Dotted lines indicate comparisons between subsets sampled from databases (ANI-1, QM9, GDB-13) and the full chemical space for that given database for QM9 and ANI-1. Note that for GDB-13, the lines do not extend across the full range of sampling fractions. Marker shapes distinguish datasets.}
    \label{fig:dbs}
\end{figure*}

\begin{table}
\centering
\begin{ruledtabular}
\begin{tabular}{lccrr}
\textbf{Dataset} & \textbf{DB size} & \textbf{Chemical space size} & \textbf{KS} & \textbf{KL} \\
\hline
ANI-1  & $5.75 \times 10^{4}$     & $3.663 \times 10^{6}$     & 0.260 & 1.291  \\
QM9    & $1.31 \times 10^{5}$     & $5.897 \times 10^{7}$     & 0.485 & 5.141  \\
GDB-13 & $9.77 \times 10^{8}$     & $1.100 \times 10^{15}$    & 0.436 & 16.389 \\
\end{tabular}
\end{ruledtabular}
\caption{Number of molecules in the original databases (DB size) and corresponding estimated chemical space sizes, along with Kolmogorov–Smirnov (KS) and Kullback–Leibler (KL) divergence scores.}
\label{tab:db_space_ks_kl}
\end{table}

Given that large-scale databases of quantum chemistry data are systematically generated in a high-throughput setting, it is just natural that some artifacts of the tooling involved affect the exhaustiveness of the result. For example, the widely used QM9 database \cite{Ramakrishnan2014} which is built to almost exhaustively contain molecules from C, H, N, O, F with $\le 9$ heavy atoms, does not contain benzene, carbon dioxide, ethylene, butadiene and other common compounds, even though they would belong to the target space of the database. 

With our method we can estimate whether the given database is likely to be representative of the underlying chemical space by comparing the distribution of molecules from the underlying chemical space with the distribution of the database entries. We performed this comparison for three related molecular databases: ANI-1 \cite{Smith2017}, QM9 \cite{Ramakrishnan2014} and GDB-13 \cite{Blum2009c}. To ensure the comparison, we first recreated the chemical space corresponding to each database using our method, replicating the main compositional constraints regarding elements and atom counts. The specific conditions for each database are: ANI-1 (H, C, N, O; $\le$ 8 heavy atoms), QM9 (H, C, N, O, F; $\le$9 heavy atoms), and GDB-13 (H, C, N, O, S, Cl; $\le$ 13 heavy atoms). GDB-13 has additional restrictions on stoichiometries and graph motifs to bias towards specific application domains which we explicitly did not consider here, as we aim to compare to less biased chemical spaces for novelty discovery.

All molecules from the database were grouped according to their chemical formula, resulting in frequency distributions of each stoichiometry. Sorting the stoichiometries according to their size in the reference chemical space allows to calculate a well-defined cumulative count of compounds as in Fig.~\ref{fig:exact-base}B. Since databases are meant to form a subset, they are naturally smaller in total count than the underlying space. To account for this, the cumulative counts where normalized by dividing the cumulative counts by the total number of compounds in either space or the database. This transforms the cumulative counts into cumulative distribution functions (CDF). 

Since CDF are a standard way to represent distributions, a wide array to comparison metrics exist. We evaluated the Kolmogorov–Smirnov (KS) test\cite{Massey1951} and the Kullback-Leibler Divergence (KL)\cite{Kullback1951}, as they have found applications in related works\cite{Rassokhin2000,Ji2022,McClendon2012a}. The KS statistic is a non-parametric test of the equality of continuous, one-dimensional probability distribution. It measures the maximum distance between the CDF of two samples. Here, those are the chemical space and the database. A lower KS value indicates that the two distributions are closely aligned over the full support, which zero being the ideal value for identical samples. The $\textrm{KL}(P\parallel Q)$ divergence instead shows how one probability distribution diverges from a second distribution.

\begin{align}
    \textrm{KL}=\sum _{x\in {\mathcal {X}}}P(x)\,\log {\frac {P(x)}{Q(x)}}\label{eqn:kldiv}
\end{align}

This asymmetric metric captures how much information is lost when using $Q$ to approximate $P$, where $P$ represents the chemical space and $Q$ the database distribution. We found KS and KL to agree in their assessment of the similarity of the databases and their underlying chemical space, but show results for KS only. This is because the KL metric contains undefined terms if one stoichiometry $x$ from the chemical space $\mathcal {X}$ is not present in the database, since then $Q(x)=0$ in eqn.~\ref{eqn:kldiv}, which in practice requires assuming some vanishingly small value for $Q(x)$ which then inflates the metric and makes it harder to compare between different databases of different chemical spaces. KS does not suffer from this issue and due to its normalisation to the interval [0,1] is easier to interpret.

Fig.~\ref{fig:dbs} shows a) how close existing databases are to their underlying chemical space (see also Tab.~\ref{tab:db_space_ks_kl} and b) which sampling fraction is a lower bound to representatively capture the distribution of stoichiometry diversity in a database. It is not only helpful to see how well a database represents the underlying target space, but also by how much a database could be further reduced while still achieving the same representativeness compared to the full database. Smaller databases would allow for higher quality reference calculations and therefore are generally more desirable. We achieve this by random subsampling of the databases in question.

For the QM9 and GDB13 datasets, the KS values are higher than for ANI-1. This behaviour is expected given the broader chemical diversity and larger size of these datasets. Higher KS scores reflect the inclusion of valid stoichiometries that are not represented in the original datasets. The small difference between QM9 and GDB13  is the consequence of them being related: QM9 is obtained by further filtering GDB13 to those molecules which converged to a DFT minimum using a particular workflow, so naturally some of the worse score of QM9 can be explained by the selection process of which molecular graphs are stable. As discussed above, this only explains part of the QM9 score as that database is also missing practically relevant molecules. KL scores allow no direct comparison between databases of different chemical spaces as the KL divergence is particularly  sensitive to mismatches in the tails of the distributions\cite{Ji2022}, which affects non-sampled low-probability stoichiometries. These outliers increase the referent distance without necessarily compromising the representativeness of the core distribution. 

Interestingly, the generated chemical spaces required significantly smaller subset sizes to approximate the overall distribution with high fidelity. As a result, the generated spaces can be characterized more efficiently with fewer samples, a useful property for downstream applications such as virtual screening \cite{Lagarde2015} or training data selection \cite{Hossain2025}.

\section{Conclusions}
Whenever we make data-driven statements in an attempt to characterize global behavior of chemical space such as structure property relationships, then the fundamental question is whether those are valid for single regions in chemical space or whether this has general predictive power. Either of them would be useful, but it is important to be able to distinguish the cases. Our method potentially allows to put any such data-driven statement on a more rigorous footing by sampling molecules from a given chemical space, making sure to consider all of them without rejection and then testing whether a statement is correct on the sampled molecules. Potentially, this could either uncover a bias in existing databases or generative model output and analysis or prove the lack thereof. In both cases, this enables work towards improved  understanding of chemical space as a whole, moving away from considering constituent molecules on a case by case basis. Statistical analysis of the generated molecules may help to identify additional trends and structural properties of chemical space. 

One of the main advantages of machine learning is the existence of formal guarantees on how a model must behave or must improve if additional data points are provided. These formal guarantees, however, usually require a certain well-behavedness of the underlying target function, such as independent and identically distributed training data. It is not clear whether current databases fulfil that requirement since ensuring a certain distribution of data points is rarely enforced in database generation. Coverage of chemical space becomes more important with multi-level machine learning methods and with our approach, relevant coverage becomes computationally feasible even for large spaces.

The main limitations of our approach lie in a) the need for a pre-built database of average path lengths, even it is shared amongst all researchers and b) the approximative nature of the exploited correlations which renders the resulting samples only approximately uniformly distributed over chemical space. Finally, some of the resulting graphs will not admit a stable minimum energy geometry; future work may address this by introducing additional weights for stoichiometries.

Finally, our work helps to address a more subtle effect regarding machine learning databases: many of them are quite old (more than ten years\cite{Ramakrishnan2014}) but are still used for benchmark. This means that over time there is a risk that we do not  obtain new models that are of lower generalization error, but rather models which, if trained on these databases, can predict the same benchmark databases well. This is a necessary consequence of using the same benchmark databases for many iterations of different models, even if each of which is designed and cross-validated following best practices. This means that in order to test generalization error, we also need test data to be a moving target as guaranteed to be unseen data\cite{Antoniuk2025}. Using the randomized sampling approach, this can be achieved easily because the space to sample from is virtually unlimited. In this approach, one could in regular intervals obtain small sets of unseen data that then affords a more truthful approximation to the generalization error in quantum chemistry. 

\bibliography{nablachem}

\begin{thebibliography}{83}%
\makeatletter
\providecommand \@ifxundefined [1]{%
 \@ifx{#1\undefined}
}%
\providecommand \@ifnum [1]{%
 \ifnum #1\expandafter \@firstoftwo
 \else \expandafter \@secondoftwo
 \fi
}%
\providecommand \@ifx [1]{%
 \ifx #1\expandafter \@firstoftwo
 \else \expandafter \@secondoftwo
 \fi
}%
\providecommand \natexlab [1]{#1}%
\providecommand \enquote  [1]{``#1''}%
\providecommand \bibnamefont  [1]{#1}%
\providecommand \bibfnamefont [1]{#1}%
\providecommand \citenamefont [1]{#1}%
\providecommand \href@noop [0]{\@secondoftwo}%
\providecommand \href [0]{\begingroup \@sanitize@url \@href}%
\providecommand \@href[1]{\@@startlink{#1}\@@href}%
\providecommand \@@href[1]{\endgroup#1\@@endlink}%
\providecommand \@sanitize@url [0]{\catcode `\\12\catcode `\$12\catcode `\&12\catcode `\#12\catcode `\^12\catcode `\_12\catcode `\%12\relax}%
\providecommand \@@startlink[1]{}%
\providecommand \@@endlink[0]{}%
\providecommand \url  [0]{\begingroup\@sanitize@url \@url }%
\providecommand \@url [1]{\endgroup\@href {#1}{\urlprefix }}%
\providecommand \urlprefix  [0]{URL }%
\providecommand \Eprint [0]{\href }%
\providecommand \doibase [0]{https://doi.org/}%
\providecommand \selectlanguage [0]{\@gobble}%
\providecommand \bibinfo  [0]{\@secondoftwo}%
\providecommand \bibfield  [0]{\@secondoftwo}%
\providecommand \translation [1]{[#1]}%
\providecommand \BibitemOpen [0]{}%
\providecommand \bibitemStop [0]{}%
\providecommand \bibitemNoStop [0]{.\EOS\space}%
\providecommand \EOS [0]{\spacefactor3000\relax}%
\providecommand \BibitemShut  [1]{\csname bibitem#1\endcsname}%
\let\auto@bib@innerbib\@empty
\bibitem [{\citenamefont {Reymond}(2015)}]{Reymond2015}%
  \BibitemOpen
  \bibfield  {author} {\bibinfo {author} {\bibfnamefont {J.-L.}\ \bibnamefont {Reymond}},\ }\bibfield  {title} {\bibinfo {title} {The {Chemical} {Space} {Project}},\ }\href {https://doi.org/10.1021/ar500432k} {\bibfield  {journal} {\bibinfo  {journal} {Accounts of Chemical Research}\ }\textbf {\bibinfo {volume} {48}},\ \bibinfo {pages} {722} (\bibinfo {year} {2015})}\BibitemShut {NoStop}%
\bibitem [{\citenamefont {Montavon}\ \emph {et~al.}(2013)\citenamefont {Montavon}, \citenamefont {Rupp}, \citenamefont {Gobre}, \citenamefont {Vazquez-Mayagoitia}, \citenamefont {Hansen}, \citenamefont {Tkatchenko}, \citenamefont {Müller},\ and\ \citenamefont {von Lilienfeld}}]{Montavon2013}%
  \BibitemOpen
  \bibfield  {author} {\bibinfo {author} {\bibfnamefont {G.}~\bibnamefont {Montavon}}, \bibinfo {author} {\bibfnamefont {M.}~\bibnamefont {Rupp}}, \bibinfo {author} {\bibfnamefont {V.}~\bibnamefont {Gobre}}, \bibinfo {author} {\bibfnamefont {A.}~\bibnamefont {Vazquez-Mayagoitia}}, \bibinfo {author} {\bibfnamefont {K.}~\bibnamefont {Hansen}}, \bibinfo {author} {\bibfnamefont {A.}~\bibnamefont {Tkatchenko}}, \bibinfo {author} {\bibfnamefont {K.-R.}\ \bibnamefont {Müller}},\ and\ \bibinfo {author} {\bibfnamefont {O.~A.}\ \bibnamefont {von Lilienfeld}},\ }\bibfield  {title} {\bibinfo {title} {Machine learning of molecular electronic properties in chemical compound space},\ }\href {https://doi.org/10.1088/1367-2630/15/9/095003} {\bibfield  {journal} {\bibinfo  {journal} {New Journal of Physics}\ }\textbf {\bibinfo {volume} {15}},\ \bibinfo {pages} {095003} (\bibinfo {year} {2013})},\ \bibinfo {note} {publisher: IOP Publishing tex.timestamp: 2019-05-12}\BibitemShut {NoStop}%
\bibitem [{\citenamefont {Keith}\ \emph {et~al.}(2021)\citenamefont {Keith}, \citenamefont {Vassilev-Galindo}, \citenamefont {Cheng}, \citenamefont {Chmiela}, \citenamefont {Gastegger}, \citenamefont {Müller},\ and\ \citenamefont {Tkatchenko}}]{keith2021combining}%
  \BibitemOpen
  \bibfield  {author} {\bibinfo {author} {\bibfnamefont {J.~A.}\ \bibnamefont {Keith}}, \bibinfo {author} {\bibfnamefont {V.}~\bibnamefont {Vassilev-Galindo}}, \bibinfo {author} {\bibfnamefont {B.}~\bibnamefont {Cheng}}, \bibinfo {author} {\bibfnamefont {S.}~\bibnamefont {Chmiela}}, \bibinfo {author} {\bibfnamefont {M.}~\bibnamefont {Gastegger}}, \bibinfo {author} {\bibfnamefont {K.-R.}\ \bibnamefont {Müller}},\ and\ \bibinfo {author} {\bibfnamefont {A.}~\bibnamefont {Tkatchenko}},\ }\bibfield  {title} {\bibinfo {title} {Combining machine learning and computational chemistry for predictive modeling and design},\ }\href {https://doi.org/10.1021/acs.chemrev.1c00107} {\bibfield  {journal} {\bibinfo  {journal} {Chemical Reviews}\ }\textbf {\bibinfo {volume} {121}},\ \bibinfo {pages} {9816} (\bibinfo {year} {2021})},\ \bibinfo {note} {publisher: American Chemical Society}\BibitemShut {NoStop}%
\bibitem [{\citenamefont {Medina‐Franco}\ \emph {et~al.}(2022)\citenamefont {Medina‐Franco}, \citenamefont {Chávez‐Hernández}, \citenamefont {López‐López},\ and\ \citenamefont {Saldívar‐González}}]{Medina-Franco2022}%
  \BibitemOpen
  \bibfield  {author} {\bibinfo {author} {\bibfnamefont {J.~L.}\ \bibnamefont {Medina‐Franco}}, \bibinfo {author} {\bibfnamefont {A.~L.}\ \bibnamefont {Chávez‐Hernández}}, \bibinfo {author} {\bibfnamefont {E.}~\bibnamefont {López‐López}},\ and\ \bibinfo {author} {\bibfnamefont {F.~I.}\ \bibnamefont {Saldívar‐González}},\ }\bibfield  {title} {\bibinfo {title} {Chemical {Multiverse}: {An} {Expanded} {View} of {Chemical} {Space}},\ }\href {https://doi.org/10.1002/minf.202200116} {\bibfield  {journal} {\bibinfo  {journal} {Molecular Informatics}\ }\textbf {\bibinfo {volume} {41}},\ \bibinfo {pages} {2200116} (\bibinfo {year} {2022})}\BibitemShut {NoStop}%
\bibitem [{\citenamefont {Hoja}\ \emph {et~al.}(2021)\citenamefont {Hoja}, \citenamefont {Medrano~Sandonas}, \citenamefont {Ernst}, \citenamefont {Vazquez-Mayagoitia}, \citenamefont {DiStasio},\ and\ \citenamefont {Tkatchenko}}]{Hoja2021}%
  \BibitemOpen
  \bibfield  {author} {\bibinfo {author} {\bibfnamefont {J.}~\bibnamefont {Hoja}}, \bibinfo {author} {\bibfnamefont {L.}~\bibnamefont {Medrano~Sandonas}}, \bibinfo {author} {\bibfnamefont {B.~G.}\ \bibnamefont {Ernst}}, \bibinfo {author} {\bibfnamefont {A.}~\bibnamefont {Vazquez-Mayagoitia}}, \bibinfo {author} {\bibfnamefont {R.~A.}\ \bibnamefont {DiStasio}},\ and\ \bibinfo {author} {\bibfnamefont {A.}~\bibnamefont {Tkatchenko}},\ }\bibfield  {title} {\bibinfo {title} {{QM7}-{X}, a comprehensive dataset of quantum-mechanical properties spanning the chemical space of small organic molecules},\ }\href {https://doi.org/10.1038/s41597-021-00812-2} {\bibfield  {journal} {\bibinfo  {journal} {Scientific Data}\ }\textbf {\bibinfo {volume} {8}},\ \bibinfo {pages} {43} (\bibinfo {year} {2021})}\BibitemShut {NoStop}%
\bibitem [{\citenamefont {Ganscha}\ \emph {et~al.}(2025)\citenamefont {Ganscha}, \citenamefont {Unke}, \citenamefont {Ahlin}, \citenamefont {Maennel}, \citenamefont {Kashubin},\ and\ \citenamefont {Müller}}]{Ganscha2025}%
  \BibitemOpen
  \bibfield  {author} {\bibinfo {author} {\bibfnamefont {S.}~\bibnamefont {Ganscha}}, \bibinfo {author} {\bibfnamefont {O.~T.}\ \bibnamefont {Unke}}, \bibinfo {author} {\bibfnamefont {D.}~\bibnamefont {Ahlin}}, \bibinfo {author} {\bibfnamefont {H.}~\bibnamefont {Maennel}}, \bibinfo {author} {\bibfnamefont {S.}~\bibnamefont {Kashubin}},\ and\ \bibinfo {author} {\bibfnamefont {K.-R.}\ \bibnamefont {Müller}},\ }\bibfield  {title} {\bibinfo {title} {The {QCML} dataset, {Quantum} chemistry reference data from 33.{5M} {DFT} and 14.{7B} semi-empirical calculations},\ }\href {https://doi.org/10.1038/s41597-025-04720-7} {\bibfield  {journal} {\bibinfo  {journal} {Scientific Data}\ }\textbf {\bibinfo {volume} {12}},\ \bibinfo {pages} {406} (\bibinfo {year} {2025})}\BibitemShut {NoStop}%
\bibitem [{\citenamefont {Smith}\ \emph {et~al.}(2017{\natexlab{a}})\citenamefont {Smith}, \citenamefont {Isayev},\ and\ \citenamefont {Roitberg}}]{Smith2017a}%
  \BibitemOpen
  \bibfield  {author} {\bibinfo {author} {\bibfnamefont {J.~S.}\ \bibnamefont {Smith}}, \bibinfo {author} {\bibfnamefont {O.}~\bibnamefont {Isayev}},\ and\ \bibinfo {author} {\bibfnamefont {A.~E.}\ \bibnamefont {Roitberg}},\ }\bibfield  {title} {\bibinfo {title} {{ANI}-1, {A} data set of 20 million calculated off-equilibrium conformations for organic molecules},\ }\href {https://doi.org/10.1038/sdata.2017.193} {\bibfield  {journal} {\bibinfo  {journal} {Scientific Data}\ }\textbf {\bibinfo {volume} {4}},\ \bibinfo {pages} {170193} (\bibinfo {year} {2017}{\natexlab{a}})}\BibitemShut {NoStop}%
\bibitem [{\citenamefont {Smith}\ \emph {et~al.}(2020)\citenamefont {Smith}, \citenamefont {Zubatyuk}, \citenamefont {Nebgen}, \citenamefont {Lubbers}, \citenamefont {Barros}, \citenamefont {Roitberg}, \citenamefont {Isayev},\ and\ \citenamefont {Tretiak}}]{Smith2020}%
  \BibitemOpen
  \bibfield  {author} {\bibinfo {author} {\bibfnamefont {J.~S.}\ \bibnamefont {Smith}}, \bibinfo {author} {\bibfnamefont {R.}~\bibnamefont {Zubatyuk}}, \bibinfo {author} {\bibfnamefont {B.}~\bibnamefont {Nebgen}}, \bibinfo {author} {\bibfnamefont {N.}~\bibnamefont {Lubbers}}, \bibinfo {author} {\bibfnamefont {K.}~\bibnamefont {Barros}}, \bibinfo {author} {\bibfnamefont {A.~E.}\ \bibnamefont {Roitberg}}, \bibinfo {author} {\bibfnamefont {O.}~\bibnamefont {Isayev}},\ and\ \bibinfo {author} {\bibfnamefont {S.}~\bibnamefont {Tretiak}},\ }\bibfield  {title} {\bibinfo {title} {The {ANI}-1ccx and {ANI}-1x data sets, coupled-cluster and density functional theory properties for molecules},\ }\bibfield  {journal} {\bibinfo  {journal} {Scientific Data}\ }\textbf {\bibinfo {volume} {7}},\ \href {https://doi.org/10.1038/s41597-020-0473-z} {10.1038/s41597-020-0473-z} (\bibinfo {year} {2020}),\ \bibinfo {note} {publisher: Springer Science and Business Media LLC}\BibitemShut {NoStop}%
\bibitem [{\citenamefont {Blum}\ and\ \citenamefont {Reymond}(2009{\natexlab{a}})}]{Blum2009}%
  \BibitemOpen
  \bibfield  {author} {\bibinfo {author} {\bibfnamefont {L.~C.}\ \bibnamefont {Blum}}\ and\ \bibinfo {author} {\bibfnamefont {J.-L.}\ \bibnamefont {Reymond}},\ }\bibfield  {title} {\bibinfo {title} {970 million druglike small molecules for virtual screening in the chemical universe database {GDB}-13},\ }\href {https://doi.org/10.1021/ja902302h} {\bibfield  {journal} {\bibinfo  {journal} {Journal of the American Chemical Society}\ }\textbf {\bibinfo {volume} {131}},\ \bibinfo {pages} {8732} (\bibinfo {year} {2009}{\natexlab{a}})},\ \bibinfo {note} {publisher: American Chemical Society (ACS) tex.timestamp: 2019-03-13}\BibitemShut {NoStop}%
\bibitem [{\citenamefont {Ruddigkeit}\ \emph {et~al.}(2012)\citenamefont {Ruddigkeit}, \citenamefont {van Deursen}, \citenamefont {Blum},\ and\ \citenamefont {Reymond}}]{Ruddigkeit2012}%
  \BibitemOpen
  \bibfield  {author} {\bibinfo {author} {\bibfnamefont {L.}~\bibnamefont {Ruddigkeit}}, \bibinfo {author} {\bibfnamefont {R.}~\bibnamefont {van Deursen}}, \bibinfo {author} {\bibfnamefont {L.~C.}\ \bibnamefont {Blum}},\ and\ \bibinfo {author} {\bibfnamefont {J.-L.}\ \bibnamefont {Reymond}},\ }\bibfield  {title} {\bibinfo {title} {Enumeration of 166 billion organic small molecules in the chemical universe database {GDB}-17},\ }\href {https://doi.org/10.1021/ci300415d} {\bibfield  {journal} {\bibinfo  {journal} {Journal of Chemical Information and Modeling}\ }\textbf {\bibinfo {volume} {52}},\ \bibinfo {pages} {2864} (\bibinfo {year} {2012})},\ \bibinfo {note} {publisher: American Chemical Society (ACS) tex.timestamp: 2018-12-20}\BibitemShut {NoStop}%
\bibitem [{\citenamefont {Ramakrishnan}\ \emph {et~al.}(2014)\citenamefont {Ramakrishnan}, \citenamefont {Dral}, \citenamefont {Rupp},\ and\ \citenamefont {von Lilienfeld}}]{Ramakrishnan2014}%
  \BibitemOpen
  \bibfield  {author} {\bibinfo {author} {\bibfnamefont {R.}~\bibnamefont {Ramakrishnan}}, \bibinfo {author} {\bibfnamefont {P.~O.}\ \bibnamefont {Dral}}, \bibinfo {author} {\bibfnamefont {M.}~\bibnamefont {Rupp}},\ and\ \bibinfo {author} {\bibfnamefont {O.~A.}\ \bibnamefont {von Lilienfeld}},\ }\bibfield  {title} {\bibinfo {title} {Quantum chemistry structures and properties of 134 kilo molecules},\ }\bibfield  {journal} {\bibinfo  {journal} {Scientific Data}\ }\textbf {\bibinfo {volume} {1}},\ \href {https://doi.org/10.1038/sdata.2014.22} {10.1038/sdata.2014.22} (\bibinfo {year} {2014}),\ \bibinfo {note} {publisher: Springer Nature tex.timestamp: 2019-03-13}\BibitemShut {NoStop}%
\bibitem [{\citenamefont {Zheng}\ \emph {et~al.}(2009)\citenamefont {Zheng}, \citenamefont {Zhao},\ and\ \citenamefont {Truhlar}}]{Zheng2009}%
  \BibitemOpen
  \bibfield  {author} {\bibinfo {author} {\bibfnamefont {J.}~\bibnamefont {Zheng}}, \bibinfo {author} {\bibfnamefont {Y.}~\bibnamefont {Zhao}},\ and\ \bibinfo {author} {\bibfnamefont {D.~G.}\ \bibnamefont {Truhlar}},\ }\bibfield  {title} {\bibinfo {title} {The {DBH24}/08 database and its use to assess electronic structure model chemistries for chemical reaction barrier heights},\ }\href {https://doi.org/10.1021/ct800568m} {\bibfield  {journal} {\bibinfo  {journal} {Journal of Chemical Theory and Computation}\ }\textbf {\bibinfo {volume} {5}},\ \bibinfo {pages} {808} (\bibinfo {year} {2009})},\ \bibinfo {note} {publisher: American Chemical Society (ACS) tex.timestamp: 2019-04-02}\BibitemShut {NoStop}%
\bibitem [{\citenamefont {Balcells}\ and\ \citenamefont {Skjelstad}(2020)}]{Balcells2020}%
  \BibitemOpen
  \bibfield  {author} {\bibinfo {author} {\bibfnamefont {D.}~\bibnamefont {Balcells}}\ and\ \bibinfo {author} {\bibfnamefont {B.~B.}\ \bibnamefont {Skjelstad}},\ }\bibfield  {title} {\bibinfo {title} {{tmQM} {Dataset}—{Quantum} {Geometries} and {Properties} of 86k {Transition} {Metal} {Complexes}},\ }\href {https://doi.org/10.1021/acs.jcim.0c01041} {\bibfield  {journal} {\bibinfo  {journal} {Journal of Chemical Information and Modeling}\ }\textbf {\bibinfo {volume} {60}},\ \bibinfo {pages} {6135} (\bibinfo {year} {2020})}\BibitemShut {NoStop}%
\bibitem [{\citenamefont {Nakata}\ \emph {et~al.}(2020)\citenamefont {Nakata}, \citenamefont {Shimazaki}, \citenamefont {Hashimoto},\ and\ \citenamefont {Maeda}}]{Nakata2020}%
  \BibitemOpen
  \bibfield  {author} {\bibinfo {author} {\bibfnamefont {M.}~\bibnamefont {Nakata}}, \bibinfo {author} {\bibfnamefont {T.}~\bibnamefont {Shimazaki}}, \bibinfo {author} {\bibfnamefont {M.}~\bibnamefont {Hashimoto}},\ and\ \bibinfo {author} {\bibfnamefont {T.}~\bibnamefont {Maeda}},\ }\bibfield  {title} {\bibinfo {title} {{PubChemQC} {PM6}: {Data} {Sets} of 221 {Million} {Molecules} with {Optimized} {Molecular} {Geometries} and {Electronic} {Properties}},\ }\href {https://doi.org/10.1021/acs.jcim.0c00740} {\bibfield  {journal} {\bibinfo  {journal} {Journal of Chemical Information and Modeling}\ }\textbf {\bibinfo {volume} {60}},\ \bibinfo {pages} {5891} (\bibinfo {year} {2020})}\BibitemShut {NoStop}%
\bibitem [{\citenamefont {Schreiner}\ \emph {et~al.}(2022)\citenamefont {Schreiner}, \citenamefont {Bhowmik}, \citenamefont {Vegge}, \citenamefont {Busk},\ and\ \citenamefont {Winther}}]{Schreiner2022}%
  \BibitemOpen
  \bibfield  {author} {\bibinfo {author} {\bibfnamefont {M.}~\bibnamefont {Schreiner}}, \bibinfo {author} {\bibfnamefont {A.}~\bibnamefont {Bhowmik}}, \bibinfo {author} {\bibfnamefont {T.}~\bibnamefont {Vegge}}, \bibinfo {author} {\bibfnamefont {J.}~\bibnamefont {Busk}},\ and\ \bibinfo {author} {\bibfnamefont {O.}~\bibnamefont {Winther}},\ }\bibfield  {title} {\bibinfo {title} {Transition1x - a dataset for building generalizable reactive machine learning potentials},\ }\href {https://doi.org/10.1038/s41597-022-01870-w} {\bibfield  {journal} {\bibinfo  {journal} {Scientific Data}\ }\textbf {\bibinfo {volume} {9}},\ \bibinfo {pages} {779} (\bibinfo {year} {2022})}\BibitemShut {NoStop}%
\bibitem [{\citenamefont {Zhao}\ \emph {et~al.}(2023)\citenamefont {Zhao}, \citenamefont {Vaddadi}, \citenamefont {Woulfe}, \citenamefont {Ogunfowora}, \citenamefont {Garimella}, \citenamefont {Isayev},\ and\ \citenamefont {Savoie}}]{Zhao2023}%
  \BibitemOpen
  \bibfield  {author} {\bibinfo {author} {\bibfnamefont {Q.}~\bibnamefont {Zhao}}, \bibinfo {author} {\bibfnamefont {S.~M.}\ \bibnamefont {Vaddadi}}, \bibinfo {author} {\bibfnamefont {M.}~\bibnamefont {Woulfe}}, \bibinfo {author} {\bibfnamefont {L.~A.}\ \bibnamefont {Ogunfowora}}, \bibinfo {author} {\bibfnamefont {S.~S.}\ \bibnamefont {Garimella}}, \bibinfo {author} {\bibfnamefont {O.}~\bibnamefont {Isayev}},\ and\ \bibinfo {author} {\bibfnamefont {B.~M.}\ \bibnamefont {Savoie}},\ }\bibfield  {title} {\bibinfo {title} {Comprehensive exploration of graphically defined reaction spaces},\ }\href {https://doi.org/10.1038/s41597-023-02043-z} {\bibfield  {journal} {\bibinfo  {journal} {Scientific Data}\ }\textbf {\bibinfo {volume} {10}},\ \bibinfo {pages} {145} (\bibinfo {year} {2023})}\BibitemShut {NoStop}%
\bibitem [{\citenamefont {Isert}\ \emph {et~al.}(2022)\citenamefont {Isert}, \citenamefont {Atz}, \citenamefont {Jiménez-Luna},\ and\ \citenamefont {Schneider}}]{Isert2022}%
  \BibitemOpen
  \bibfield  {author} {\bibinfo {author} {\bibfnamefont {C.}~\bibnamefont {Isert}}, \bibinfo {author} {\bibfnamefont {K.}~\bibnamefont {Atz}}, \bibinfo {author} {\bibfnamefont {J.}~\bibnamefont {Jiménez-Luna}},\ and\ \bibinfo {author} {\bibfnamefont {G.}~\bibnamefont {Schneider}},\ }\bibfield  {title} {\bibinfo {title} {{QMugs}, quantum mechanical properties of drug-like molecules},\ }\href {https://doi.org/10.1038/s41597-022-01390-7} {\bibfield  {journal} {\bibinfo  {journal} {Scientific Data}\ }\textbf {\bibinfo {volume} {9}},\ \bibinfo {pages} {273} (\bibinfo {year} {2022})}\BibitemShut {NoStop}%
\bibitem [{\citenamefont {Ehlert}\ \emph {et~al.}(2025)\citenamefont {Ehlert}, \citenamefont {Hermann}, \citenamefont {Vogels}, \citenamefont {Satorras}, \citenamefont {Lanius}, \citenamefont {Segler}, \citenamefont {Kooi}, \citenamefont {Takeda}, \citenamefont {Huang}, \citenamefont {Luise}, \citenamefont {Berg}, \citenamefont {Gori-Giorgi},\ and\ \citenamefont {Karton}}]{Ehlert2025}%
  \BibitemOpen
  \bibfield  {author} {\bibinfo {author} {\bibfnamefont {S.}~\bibnamefont {Ehlert}}, \bibinfo {author} {\bibfnamefont {J.}~\bibnamefont {Hermann}}, \bibinfo {author} {\bibfnamefont {T.}~\bibnamefont {Vogels}}, \bibinfo {author} {\bibfnamefont {V.~G.}\ \bibnamefont {Satorras}}, \bibinfo {author} {\bibfnamefont {S.}~\bibnamefont {Lanius}}, \bibinfo {author} {\bibfnamefont {M.}~\bibnamefont {Segler}}, \bibinfo {author} {\bibfnamefont {D.~P.}\ \bibnamefont {Kooi}}, \bibinfo {author} {\bibfnamefont {K.}~\bibnamefont {Takeda}}, \bibinfo {author} {\bibfnamefont {C.-W.}\ \bibnamefont {Huang}}, \bibinfo {author} {\bibfnamefont {G.}~\bibnamefont {Luise}}, \bibinfo {author} {\bibfnamefont {R.~v.~d.}\ \bibnamefont {Berg}}, \bibinfo {author} {\bibfnamefont {P.}~\bibnamefont {Gori-Giorgi}},\ and\ \bibinfo {author} {\bibfnamefont {A.}~\bibnamefont {Karton}},\ }\href {https://doi.org/10.48550/ARXIV.2506.14492} {\bibinfo {title} {Accurate {Chemistry} {Collection}: {Coupled} cluster atomization energies for broad chemical
  space}} (\bibinfo {year} {2025}),\ \bibinfo {note} {version Number: 3}\BibitemShut {NoStop}%
\bibitem [{\citenamefont {Chakraborty}\ \emph {et~al.}(2025)\citenamefont {Chakraborty}, \citenamefont {Almog},\ and\ \citenamefont {Gershoni-Poranne}}]{Chakraborty2025}%
  \BibitemOpen
  \bibfield  {author} {\bibinfo {author} {\bibfnamefont {S.}~\bibnamefont {Chakraborty}}, \bibinfo {author} {\bibfnamefont {I.}~\bibnamefont {Almog}},\ and\ \bibinfo {author} {\bibfnamefont {R.}~\bibnamefont {Gershoni-Poranne}},\ }\bibfield  {title} {\bibinfo {title} {{COMPAS}-4: {A} {Data} {Set} of ({BN})$_{\textrm{1}}$ {Substituted} \textit{{Cata}} -{Condensed} {Polybenzenoid} {Hydrocarbons}-{Data} {Analysis} and {Feature} {Engineering}},\ }\href {https://doi.org/10.1021/acs.jcim.5c00608} {\bibfield  {journal} {\bibinfo  {journal} {Journal of Chemical Information and Modeling}\ }\textbf {\bibinfo {volume} {65}},\ \bibinfo {pages} {5508} (\bibinfo {year} {2025})}\BibitemShut {NoStop}%
\bibitem [{\citenamefont {Ullah}\ \emph {et~al.}(2024)\citenamefont {Ullah}, \citenamefont {Chen},\ and\ \citenamefont {Dral}}]{Ullah2024}%
  \BibitemOpen
  \bibfield  {author} {\bibinfo {author} {\bibfnamefont {A.}~\bibnamefont {Ullah}}, \bibinfo {author} {\bibfnamefont {Y.}~\bibnamefont {Chen}},\ and\ \bibinfo {author} {\bibfnamefont {P.~O.}\ \bibnamefont {Dral}},\ }\bibfield  {title} {\bibinfo {title} {Molecular quantum chemical data sets and databases for machine learning potentials},\ }\href {https://doi.org/10.1088/2632-2153/ad8f13} {\bibfield  {journal} {\bibinfo  {journal} {Machine Learning: Science and Technology}\ }\textbf {\bibinfo {volume} {5}},\ \bibinfo {pages} {041001} (\bibinfo {year} {2024})}\BibitemShut {NoStop}%
\bibitem [{\citenamefont {Levine}\ \emph {et~al.}(2025)\citenamefont {Levine}, \citenamefont {Shuaibi}, \citenamefont {Spotte-Smith}, \citenamefont {Taylor}, \citenamefont {Hasyim}, \citenamefont {Michel}, \citenamefont {Batatia}, \citenamefont {Csányi}, \citenamefont {Dzamba}, \citenamefont {Eastman}, \citenamefont {Frey}, \citenamefont {Fu}, \citenamefont {Gharakhanyan}, \citenamefont {Krishnapriyan}, \citenamefont {Rackers}, \citenamefont {Raja}, \citenamefont {Rizvi}, \citenamefont {Rosen}, \citenamefont {Ulissi}, \citenamefont {Vargas}, \citenamefont {Zitnick}, \citenamefont {Blau},\ and\ \citenamefont {Wood}}]{Levine2025}%
  \BibitemOpen
  \bibfield  {author} {\bibinfo {author} {\bibfnamefont {D.~S.}\ \bibnamefont {Levine}}, \bibinfo {author} {\bibfnamefont {M.}~\bibnamefont {Shuaibi}}, \bibinfo {author} {\bibfnamefont {E.~W.~C.}\ \bibnamefont {Spotte-Smith}}, \bibinfo {author} {\bibfnamefont {M.~G.}\ \bibnamefont {Taylor}}, \bibinfo {author} {\bibfnamefont {M.~R.}\ \bibnamefont {Hasyim}}, \bibinfo {author} {\bibfnamefont {K.}~\bibnamefont {Michel}}, \bibinfo {author} {\bibfnamefont {I.}~\bibnamefont {Batatia}}, \bibinfo {author} {\bibfnamefont {G.}~\bibnamefont {Csányi}}, \bibinfo {author} {\bibfnamefont {M.}~\bibnamefont {Dzamba}}, \bibinfo {author} {\bibfnamefont {P.}~\bibnamefont {Eastman}}, \bibinfo {author} {\bibfnamefont {N.~C.}\ \bibnamefont {Frey}}, \bibinfo {author} {\bibfnamefont {X.}~\bibnamefont {Fu}}, \bibinfo {author} {\bibfnamefont {V.}~\bibnamefont {Gharakhanyan}}, \bibinfo {author} {\bibfnamefont {A.~S.}\ \bibnamefont {Krishnapriyan}}, \bibinfo {author} {\bibfnamefont {J.~A.}\ \bibnamefont {Rackers}}, \bibinfo {author}
  {\bibfnamefont {S.}~\bibnamefont {Raja}}, \bibinfo {author} {\bibfnamefont {A.}~\bibnamefont {Rizvi}}, \bibinfo {author} {\bibfnamefont {A.~S.}\ \bibnamefont {Rosen}}, \bibinfo {author} {\bibfnamefont {Z.}~\bibnamefont {Ulissi}}, \bibinfo {author} {\bibfnamefont {S.}~\bibnamefont {Vargas}}, \bibinfo {author} {\bibfnamefont {C.~L.}\ \bibnamefont {Zitnick}}, \bibinfo {author} {\bibfnamefont {S.~M.}\ \bibnamefont {Blau}},\ and\ \bibinfo {author} {\bibfnamefont {B.~M.}\ \bibnamefont {Wood}},\ }\href {https://doi.org/10.48550/ARXIV.2505.08762} {\bibinfo {title} {The {Open} {Molecules} 2025 ({OMol25}) {Dataset}, {Evaluations}, and {Models}}} (\bibinfo {year} {2025}),\ \bibinfo {note} {version Number: 1}\BibitemShut {NoStop}%
\bibitem [{\citenamefont {Hossain}\ \emph {et~al.}(2025)\citenamefont {Hossain}, \citenamefont {Thiagarajan}, \citenamefont {Pathrudkar}, \citenamefont {Taylor}, \citenamefont {Gangan}, \citenamefont {Banerjee},\ and\ \citenamefont {Ghosh}}]{Hossain2025}%
  \BibitemOpen
  \bibfield  {author} {\bibinfo {author} {\bibfnamefont {S.}~\bibnamefont {Hossain}}, \bibinfo {author} {\bibfnamefont {P.}~\bibnamefont {Thiagarajan}}, \bibinfo {author} {\bibfnamefont {S.}~\bibnamefont {Pathrudkar}}, \bibinfo {author} {\bibfnamefont {S.}~\bibnamefont {Taylor}}, \bibinfo {author} {\bibfnamefont {A.~S.}\ \bibnamefont {Gangan}}, \bibinfo {author} {\bibfnamefont {A.~S.}\ \bibnamefont {Banerjee}},\ and\ \bibinfo {author} {\bibfnamefont {S.}~\bibnamefont {Ghosh}},\ }\href {https://doi.org/10.48550/ARXIV.2507.09001} {\bibinfo {title} {Surprisingly {High} {Redundancy} in {Electronic} {Structure} {Data}}} (\bibinfo {year} {2025}),\ \bibinfo {note} {version Number: 1}\BibitemShut {NoStop}%
\bibitem [{\citenamefont {Lopez~Perez}\ \emph {et~al.}(2025)\citenamefont {Lopez~Perez}, \citenamefont {López-López}, \citenamefont {Soulage}, \citenamefont {Felix}, \citenamefont {Medina-Franco},\ and\ \citenamefont {Miranda-Quintana}}]{LopezPerez2025}%
  \BibitemOpen
  \bibfield  {author} {\bibinfo {author} {\bibfnamefont {K.}~\bibnamefont {Lopez~Perez}}, \bibinfo {author} {\bibfnamefont {E.}~\bibnamefont {López-López}}, \bibinfo {author} {\bibfnamefont {F.}~\bibnamefont {Soulage}}, \bibinfo {author} {\bibfnamefont {E.}~\bibnamefont {Felix}}, \bibinfo {author} {\bibfnamefont {J.~L.}\ \bibnamefont {Medina-Franco}},\ and\ \bibinfo {author} {\bibfnamefont {R.~A.}\ \bibnamefont {Miranda-Quintana}},\ }\bibfield  {title} {\bibinfo {title} {Growth vs {Diversity}: {A} {Time}-{Evolution} {Analysis} of the {Chemical} {Space}},\ }\href {https://doi.org/10.1021/acs.jcim.5c00347} {\bibfield  {journal} {\bibinfo  {journal} {Journal of Chemical Information and Modeling}\ }\textbf {\bibinfo {volume} {65}},\ \bibinfo {pages} {6788} (\bibinfo {year} {2025})}\BibitemShut {NoStop}%
\bibitem [{\citenamefont {Duan}\ \emph {et~al.}(2019{\natexlab{a}})\citenamefont {Duan}, \citenamefont {Janet}, \citenamefont {Liu}, \citenamefont {Nandy},\ and\ \citenamefont {Kulik}}]{Duan2019}%
  \BibitemOpen
  \bibfield  {author} {\bibinfo {author} {\bibfnamefont {C.}~\bibnamefont {Duan}}, \bibinfo {author} {\bibfnamefont {J.~P.}\ \bibnamefont {Janet}}, \bibinfo {author} {\bibfnamefont {F.}~\bibnamefont {Liu}}, \bibinfo {author} {\bibfnamefont {A.}~\bibnamefont {Nandy}},\ and\ \bibinfo {author} {\bibfnamefont {H.~J.}\ \bibnamefont {Kulik}},\ }\bibfield  {title} {\bibinfo {title} {Learning from failure: {Predicting} electronic structure calculation outcomes with machine learning models},\ }\href {https://doi.org/10.1021/acs.jctc.9b00057} {\bibfield  {journal} {\bibinfo  {journal} {Journal of Chemical Theory and Computation}\ }\textbf {\bibinfo {volume} {15}},\ \bibinfo {pages} {2331} (\bibinfo {year} {2019}{\natexlab{a}})},\ \bibinfo {note} {publisher: American Chemical Society (ACS)}\BibitemShut {NoStop}%
\bibitem [{\citenamefont {Heinen}\ \emph {et~al.}(2019)\citenamefont {Heinen}, \citenamefont {Schwilk}, \citenamefont {von Rudorff},\ and\ \citenamefont {von Lilienfeld}}]{Heinen_2020}%
  \BibitemOpen
  \bibfield  {author} {\bibinfo {author} {\bibfnamefont {S.}~\bibnamefont {Heinen}}, \bibinfo {author} {\bibfnamefont {M.}~\bibnamefont {Schwilk}}, \bibinfo {author} {\bibfnamefont {G.~F.}\ \bibnamefont {von Rudorff}},\ and\ \bibinfo {author} {\bibfnamefont {O.~A.}\ \bibnamefont {von Lilienfeld}},\ }\bibfield  {title} {\bibinfo {title} {Machine learning the computational cost of quantum chemistry},\ }\href {https://doi.org/10.1088/2632-2153/ab6ac4} {\bibfield  {journal} {\bibinfo  {journal} {Machine Learning: Science and Technology}\ }\textbf {\bibinfo {volume} {1}},\ \bibinfo {pages} {025002} (\bibinfo {year} {2019})},\ \bibinfo {note} {arXiv: http://arxiv.org/abs/1908.06714v1 [physics.chem-ph] Publisher: IOP Publishing tex.timestamp: 2019-10-22}\BibitemShut {NoStop}%
\bibitem [{\citenamefont {Hammond}(1955)}]{Hammond1955}%
  \BibitemOpen
  \bibfield  {author} {\bibinfo {author} {\bibfnamefont {G.~S.}\ \bibnamefont {Hammond}},\ }\bibfield  {title} {\bibinfo {title} {A {Correlation} of {Reaction} {Rates}},\ }\href {https://doi.org/10.1021/ja01607a027} {\bibfield  {journal} {\bibinfo  {journal} {Journal of the American Chemical Society}\ }\textbf {\bibinfo {volume} {77}},\ \bibinfo {pages} {334} (\bibinfo {year} {1955})},\ \bibinfo {note} {publisher: American Chemical Society (ACS)}\BibitemShut {NoStop}%
\bibitem [{\citenamefont {Hückel}(1931)}]{Hueckel1931}%
  \BibitemOpen
  \bibfield  {author} {\bibinfo {author} {\bibfnamefont {E.}~\bibnamefont {Hückel}},\ }\bibfield  {title} {\bibinfo {title} {Quantentheoretische {Beiträge} zum {Benzolproblem}: {I}. {Die} {Elektronenkonfiguration} des {Benzols} und verwandter {Verbindungen}},\ }\href {https://doi.org/10.1007/bf01339530} {\bibfield  {journal} {\bibinfo  {journal} {Zeitschrift für Physik}\ }\textbf {\bibinfo {volume} {70}},\ \bibinfo {pages} {204} (\bibinfo {year} {1931})},\ \bibinfo {note} {publisher: Springer Science and Business Media LLC}\BibitemShut {NoStop}%
\bibitem [{\citenamefont {Heinen}\ \emph {et~al.}(2021)\citenamefont {Heinen}, \citenamefont {von Rudorff},\ and\ \citenamefont {von Lilienfeld}}]{Heinen2021}%
  \BibitemOpen
  \bibfield  {author} {\bibinfo {author} {\bibfnamefont {S.}~\bibnamefont {Heinen}}, \bibinfo {author} {\bibfnamefont {G.~F.}\ \bibnamefont {von Rudorff}},\ and\ \bibinfo {author} {\bibfnamefont {O.~A.}\ \bibnamefont {von Lilienfeld}},\ }\bibfield  {title} {\bibinfo {title} {Toward the design of chemical reactions: {Machine} learning barriers of competing mechanisms in reactant space},\ }\href {https://doi.org/10.1063/5.0059742} {\bibfield  {journal} {\bibinfo  {journal} {The Journal of Chemical Physics}\ }\textbf {\bibinfo {volume} {155}},\ \bibinfo {pages} {064105} (\bibinfo {year} {2021})},\ \bibinfo {note} {publisher: AIP Publishing}\BibitemShut {NoStop}%
\bibitem [{\citenamefont {Arteca}\ and\ \citenamefont {Mezey}(1988)}]{Arteca1988}%
  \BibitemOpen
  \bibfield  {author} {\bibinfo {author} {\bibfnamefont {G.~A.}\ \bibnamefont {Arteca}}\ and\ \bibinfo {author} {\bibfnamefont {P.~G.}\ \bibnamefont {Mezey}},\ }\bibfield  {title} {\bibinfo {title} {Validity of the {Hammond} postulate and constraints on general one‐dimensional reaction barriers},\ }\href {https://doi.org/10.1002/jcc.540090704} {\bibfield  {journal} {\bibinfo  {journal} {Journal of Computational Chemistry}\ }\textbf {\bibinfo {volume} {9}},\ \bibinfo {pages} {728} (\bibinfo {year} {1988})},\ \bibinfo {note} {publisher: Wiley}\BibitemShut {NoStop}%
\bibitem [{\citenamefont {Donahue}(2001)}]{Donahue2001}%
  \BibitemOpen
  \bibfield  {author} {\bibinfo {author} {\bibfnamefont {N.~M.}\ \bibnamefont {Donahue}},\ }\bibfield  {title} {\bibinfo {title} {Revisiting the {Hammond} {Postulate}: {The} {Role} of {Reactant} and {Product} {Ionic} {States} in {Regulating} {Barrier} {Heights}, {Locations}, and {Transition} {State} {Frequencies}},\ }\href {https://doi.org/10.1021/jp001004t} {\bibfield  {journal} {\bibinfo  {journal} {The Journal of Physical Chemistry A}\ }\textbf {\bibinfo {volume} {105}},\ \bibinfo {pages} {1489} (\bibinfo {year} {2001})},\ \bibinfo {note} {publisher: American Chemical Society (ACS)}\BibitemShut {NoStop}%
\bibitem [{\citenamefont {Van~Nyvel}\ \emph {et~al.}(2025)\citenamefont {Van~Nyvel}, \citenamefont {Alonso},\ and\ \citenamefont {Solà}}]{VanNyvel2025}%
  \BibitemOpen
  \bibfield  {author} {\bibinfo {author} {\bibfnamefont {L.}~\bibnamefont {Van~Nyvel}}, \bibinfo {author} {\bibfnamefont {M.}~\bibnamefont {Alonso}},\ and\ \bibinfo {author} {\bibfnamefont {M.}~\bibnamefont {Solà}},\ }\bibfield  {title} {\bibinfo {title} {Effect of size, charge, and spin state on {Hückel} and {Baird} aromaticity in [\textit{{N}}]annulenes},\ }\href {https://doi.org/10.1039/d4sc08225g} {\bibfield  {journal} {\bibinfo  {journal} {Chemical Science}\ }\textbf {\bibinfo {volume} {16}},\ \bibinfo {pages} {5613} (\bibinfo {year} {2025})},\ \bibinfo {note} {publisher: Royal Society of Chemistry (RSC)}\BibitemShut {NoStop}%
\bibitem [{\citenamefont {Apriliyanto}\ \emph {et~al.}(2021)\citenamefont {Apriliyanto}, \citenamefont {Battaglia}, \citenamefont {Evangelisti}, \citenamefont {Faginas-Lago}, \citenamefont {Leininger},\ and\ \citenamefont {Lombardi}}]{Apriliyanto2021}%
  \BibitemOpen
  \bibfield  {author} {\bibinfo {author} {\bibfnamefont {Y.~B.}\ \bibnamefont {Apriliyanto}}, \bibinfo {author} {\bibfnamefont {S.}~\bibnamefont {Battaglia}}, \bibinfo {author} {\bibfnamefont {S.}~\bibnamefont {Evangelisti}}, \bibinfo {author} {\bibfnamefont {N.}~\bibnamefont {Faginas-Lago}}, \bibinfo {author} {\bibfnamefont {T.}~\bibnamefont {Leininger}},\ and\ \bibinfo {author} {\bibfnamefont {A.}~\bibnamefont {Lombardi}},\ }\bibfield  {title} {\bibinfo {title} {Toward a {Generalized} {Hückel} {Rule}: {The} {Electronic} {Structure} of {Carbon} {Nanocones}},\ }\href {https://doi.org/10.1021/acs.jpca.1c06402} {\bibfield  {journal} {\bibinfo  {journal} {The Journal of Physical Chemistry A}\ }\textbf {\bibinfo {volume} {125}},\ \bibinfo {pages} {9819} (\bibinfo {year} {2021})},\ \bibinfo {note} {publisher: American Chemical Society (ACS)}\BibitemShut {NoStop}%
\bibitem [{\citenamefont {Huang}\ \emph {et~al.}(2023)\citenamefont {Huang}, \citenamefont {von Rudorff},\ and\ \citenamefont {von Lilienfeld}}]{Huang2023}%
  \BibitemOpen
  \bibfield  {author} {\bibinfo {author} {\bibfnamefont {B.}~\bibnamefont {Huang}}, \bibinfo {author} {\bibfnamefont {G.~F.}\ \bibnamefont {von Rudorff}},\ and\ \bibinfo {author} {\bibfnamefont {O.~A.}\ \bibnamefont {von Lilienfeld}},\ }\bibfield  {title} {\bibinfo {title} {The central role of density functional theory in the {AI} age},\ }\href {https://doi.org/10.1126/science.abn3445} {\bibfield  {journal} {\bibinfo  {journal} {Science}\ }\textbf {\bibinfo {volume} {381}},\ \bibinfo {pages} {170} (\bibinfo {year} {2023})},\ \bibinfo {note} {publisher: American Association for the Advancement of Science (AAAS)}\BibitemShut {NoStop}%
\bibitem [{\citenamefont {Lavecchia}(2024)}]{Lavecchia2024}%
  \BibitemOpen
  \bibfield  {author} {\bibinfo {author} {\bibfnamefont {A.}~\bibnamefont {Lavecchia}},\ }\bibfield  {title} {\bibinfo {title} {Navigating the frontier of drug-like chemical space with cutting-edge generative {AI} models},\ }\href {https://doi.org/10.1016/j.drudis.2024.104133} {\bibfield  {journal} {\bibinfo  {journal} {Drug Discovery Today}\ }\textbf {\bibinfo {volume} {29}},\ \bibinfo {pages} {104133} (\bibinfo {year} {2024})},\ \bibinfo {note} {publisher: Elsevier BV}\BibitemShut {NoStop}%
\bibitem [{\citenamefont {Sadybekov}\ and\ \citenamefont {Katritch}(2023)}]{Sadybekov2023}%
  \BibitemOpen
  \bibfield  {author} {\bibinfo {author} {\bibfnamefont {A.~V.}\ \bibnamefont {Sadybekov}}\ and\ \bibinfo {author} {\bibfnamefont {V.}~\bibnamefont {Katritch}},\ }\bibfield  {title} {\bibinfo {title} {Computational approaches streamlining drug discovery},\ }\href {https://doi.org/10.1038/s41586-023-05905-z} {\bibfield  {journal} {\bibinfo  {journal} {Nature}\ }\textbf {\bibinfo {volume} {616}},\ \bibinfo {pages} {673} (\bibinfo {year} {2023})},\ \bibinfo {note} {publisher: Springer Science and Business Media LLC}\BibitemShut {NoStop}%
\bibitem [{\citenamefont {Griego}\ \emph {et~al.}(2021)\citenamefont {Griego}, \citenamefont {Maldonado}, \citenamefont {Zhao}, \citenamefont {Zulueta}, \citenamefont {Gentry}, \citenamefont {Lipsman}, \citenamefont {Choi},\ and\ \citenamefont {Keith}}]{Griego2021}%
  \BibitemOpen
  \bibfield  {author} {\bibinfo {author} {\bibfnamefont {C.~D.}\ \bibnamefont {Griego}}, \bibinfo {author} {\bibfnamefont {A.~M.}\ \bibnamefont {Maldonado}}, \bibinfo {author} {\bibfnamefont {L.}~\bibnamefont {Zhao}}, \bibinfo {author} {\bibfnamefont {B.}~\bibnamefont {Zulueta}}, \bibinfo {author} {\bibfnamefont {B.~M.}\ \bibnamefont {Gentry}}, \bibinfo {author} {\bibfnamefont {E.}~\bibnamefont {Lipsman}}, \bibinfo {author} {\bibfnamefont {T.~H.}\ \bibnamefont {Choi}},\ and\ \bibinfo {author} {\bibfnamefont {J.~A.}\ \bibnamefont {Keith}},\ }\bibfield  {title} {\bibinfo {title} {Computationally guided searches for efficient catalysts through chemical/materials space: {Progress} and outlook},\ }\href {https://doi.org/10.1021/acs.jpcc.0c11345} {\bibfield  {journal} {\bibinfo  {journal} {The Journal of Physical Chemistry C}\ }\textbf {\bibinfo {volume} {125}},\ \bibinfo {pages} {6495} (\bibinfo {year} {2021})},\ \bibinfo {note} {publisher: American Chemical Society (ACS)}\BibitemShut {NoStop}%
\bibitem [{\citenamefont {Tan}\ \emph {et~al.}(2025)\citenamefont {Tan}, \citenamefont {Yang},\ and\ \citenamefont {Luo}}]{Tan2025}%
  \BibitemOpen
  \bibfield  {author} {\bibinfo {author} {\bibfnamefont {Z.}~\bibnamefont {Tan}}, \bibinfo {author} {\bibfnamefont {Q.}~\bibnamefont {Yang}},\ and\ \bibinfo {author} {\bibfnamefont {S.}~\bibnamefont {Luo}},\ }\bibfield  {title} {\bibinfo {title} {{AI} molecular catalysis: where are we now?},\ }\href {https://doi.org/10.1039/d4qo02363c} {\bibfield  {journal} {\bibinfo  {journal} {Organic Chemistry Frontiers}\ }\textbf {\bibinfo {volume} {12}},\ \bibinfo {pages} {2759} (\bibinfo {year} {2025})},\ \bibinfo {note} {publisher: Royal Society of Chemistry (RSC)}\BibitemShut {NoStop}%
\bibitem [{\citenamefont {Mace}\ \emph {et~al.}(2024)\citenamefont {Mace}, \citenamefont {Xu},\ and\ \citenamefont {Nguyen}}]{Mace2024}%
  \BibitemOpen
  \bibfield  {author} {\bibinfo {author} {\bibfnamefont {S.}~\bibnamefont {Mace}}, \bibinfo {author} {\bibfnamefont {Y.}~\bibnamefont {Xu}},\ and\ \bibinfo {author} {\bibfnamefont {B.~N.}\ \bibnamefont {Nguyen}},\ }\bibfield  {title} {\bibinfo {title} {Automated {Transition} {Metal} {Catalysts} {Discovery} and {Optimisation} with {AI} and {Machine} {Learning}},\ }\bibfield  {journal} {\bibinfo  {journal} {ChemCatChem}\ }\textbf {\bibinfo {volume} {16}},\ \href {https://doi.org/10.1002/cctc.202301475} {10.1002/cctc.202301475} (\bibinfo {year} {2024}),\ \bibinfo {note} {publisher: Wiley}\BibitemShut {NoStop}%
\bibitem [{\citenamefont {Gómez-Bombarelli}\ \emph {et~al.}(2018)\citenamefont {Gómez-Bombarelli}, \citenamefont {Wei}, \citenamefont {Duvenaud}, \citenamefont {Hernández-Lobato}, \citenamefont {Sánchez-Lengeling}, \citenamefont {Sheberla}, \citenamefont {Aguilera-Iparraguirre}, \citenamefont {Hirzel}, \citenamefont {Adams},\ and\ \citenamefont {Aspuru-Guzik}}]{GomezBombarelli2018}%
  \BibitemOpen
  \bibfield  {author} {\bibinfo {author} {\bibfnamefont {R.}~\bibnamefont {Gómez-Bombarelli}}, \bibinfo {author} {\bibfnamefont {J.~N.}\ \bibnamefont {Wei}}, \bibinfo {author} {\bibfnamefont {D.}~\bibnamefont {Duvenaud}}, \bibinfo {author} {\bibfnamefont {J.~M.}\ \bibnamefont {Hernández-Lobato}}, \bibinfo {author} {\bibfnamefont {B.}~\bibnamefont {Sánchez-Lengeling}}, \bibinfo {author} {\bibfnamefont {D.}~\bibnamefont {Sheberla}}, \bibinfo {author} {\bibfnamefont {J.}~\bibnamefont {Aguilera-Iparraguirre}}, \bibinfo {author} {\bibfnamefont {T.~D.}\ \bibnamefont {Hirzel}}, \bibinfo {author} {\bibfnamefont {R.~P.}\ \bibnamefont {Adams}},\ and\ \bibinfo {author} {\bibfnamefont {A.}~\bibnamefont {Aspuru-Guzik}},\ }\bibfield  {title} {\bibinfo {title} {Automatic chemical design using a data-driven continuous representation of molecules},\ }\href {https://doi.org/10.1021/acscentsci.7b00572} {\bibfield  {journal} {\bibinfo  {journal} {ACS Central Science}\ }\textbf {\bibinfo {volume} {4}},\ \bibinfo {pages} {268}
  (\bibinfo {year} {2018})},\ \bibinfo {note} {publisher: American Chemical Society (ACS)}\BibitemShut {NoStop}%
\bibitem [{\citenamefont {Cao}\ \emph {et~al.}(2025)\citenamefont {Cao}, \citenamefont {Zhang}, \citenamefont {Sun}, \citenamefont {Yin},\ and\ \citenamefont {Feng}}]{Cao2025}%
  \BibitemOpen
  \bibfield  {author} {\bibinfo {author} {\bibfnamefont {X.}~\bibnamefont {Cao}}, \bibinfo {author} {\bibfnamefont {Y.}~\bibnamefont {Zhang}}, \bibinfo {author} {\bibfnamefont {Z.}~\bibnamefont {Sun}}, \bibinfo {author} {\bibfnamefont {H.}~\bibnamefont {Yin}},\ and\ \bibinfo {author} {\bibfnamefont {Y.}~\bibnamefont {Feng}},\ }\bibfield  {title} {\bibinfo {title} {Machine learning in polymer science: {A} new lens for physical and chemical exploration},\ }\href {https://doi.org/10.1016/j.pmatsci.2025.101544} {\bibfield  {journal} {\bibinfo  {journal} {Progress in Materials Science}\ ,\ \bibinfo {pages} {101544}} (\bibinfo {year} {2025})}\BibitemShut {NoStop}%
\bibitem [{\citenamefont {Liu}\ and\ \citenamefont {Li}(2023)}]{Liu2023}%
  \BibitemOpen
  \bibfield  {author} {\bibinfo {author} {\bibfnamefont {Y.}~\bibnamefont {Liu}}\ and\ \bibinfo {author} {\bibfnamefont {Z.}~\bibnamefont {Li}},\ }\bibfield  {title} {\bibinfo {title} {Predict {Ionization} {Energy} of {Molecules} {Using} {Conventional} and {Graph}-{Based} {Machine} {Learning} {Models}},\ }\href {https://doi.org/10.1021/acs.jcim.2c01321} {\bibfield  {journal} {\bibinfo  {journal} {Journal of Chemical Information and Modeling}\ }\textbf {\bibinfo {volume} {63}},\ \bibinfo {pages} {806} (\bibinfo {year} {2023})}\BibitemShut {NoStop}%
\bibitem [{\citenamefont {Yang}\ \emph {et~al.}(2025)\citenamefont {Yang}, \citenamefont {Song}, \citenamefont {Cheng},\ and\ \citenamefont {Ren}}]{Yang2025}%
  \BibitemOpen
  \bibfield  {author} {\bibinfo {author} {\bibfnamefont {M.}~\bibnamefont {Yang}}, \bibinfo {author} {\bibfnamefont {G.}~\bibnamefont {Song}}, \bibinfo {author} {\bibfnamefont {L.}~\bibnamefont {Cheng}},\ and\ \bibinfo {author} {\bibfnamefont {H.}~\bibnamefont {Ren}},\ }\bibfield  {title} {\bibinfo {title} {{SMILES} {Token} {Additivity} {Model} with {Interpretability} and {Generalizability} for {Fuel} {Property} {Predictions}},\ }\href {https://doi.org/10.1021/acs.jcim.5c00986} {\bibfield  {journal} {\bibinfo  {journal} {Journal of Chemical Information and Modeling}\ ,\ \bibinfo {pages} {acs.jcim.5c00986}} (\bibinfo {year} {2025})}\BibitemShut {NoStop}%
\bibitem [{\citenamefont {Sun}\ \emph {et~al.}(2025)\citenamefont {Sun}, \citenamefont {Cao}, \citenamefont {Hu},\ and\ \citenamefont {Qi}}]{Sun2025a}%
  \BibitemOpen
  \bibfield  {author} {\bibinfo {author} {\bibfnamefont {J.}~\bibnamefont {Sun}}, \bibinfo {author} {\bibfnamefont {Y.}~\bibnamefont {Cao}}, \bibinfo {author} {\bibfnamefont {H.}~\bibnamefont {Hu}},\ and\ \bibinfo {author} {\bibfnamefont {B.}~\bibnamefont {Qi}},\ }\bibfield  {title} {\bibinfo {title} {Equivariant learning leveraging geometric invariances in {3D} molecular conformers for accurate prediction of quantum chemical properties},\ }\bibfield  {journal} {\bibinfo  {journal} {Scientific Reports}\ }\textbf {\bibinfo {volume} {15}},\ \href {https://doi.org/10.1038/s41598-025-09842-x} {10.1038/s41598-025-09842-x} (\bibinfo {year} {2025}),\ \bibinfo {note} {publisher: Springer Science and Business Media LLC}\BibitemShut {NoStop}%
\bibitem [{\citenamefont {Suman}\ \emph {et~al.}(2025)\citenamefont {Suman}, \citenamefont {Nigam}, \citenamefont {Saade}, \citenamefont {Pegolo}, \citenamefont {Türk}, \citenamefont {Zhang}, \citenamefont {Chan},\ and\ \citenamefont {Ceriotti}}]{Suman2025}%
  \BibitemOpen
  \bibfield  {author} {\bibinfo {author} {\bibfnamefont {D.}~\bibnamefont {Suman}}, \bibinfo {author} {\bibfnamefont {J.}~\bibnamefont {Nigam}}, \bibinfo {author} {\bibfnamefont {S.}~\bibnamefont {Saade}}, \bibinfo {author} {\bibfnamefont {P.}~\bibnamefont {Pegolo}}, \bibinfo {author} {\bibfnamefont {H.}~\bibnamefont {Türk}}, \bibinfo {author} {\bibfnamefont {X.}~\bibnamefont {Zhang}}, \bibinfo {author} {\bibfnamefont {G.~K.-L.}\ \bibnamefont {Chan}},\ and\ \bibinfo {author} {\bibfnamefont {M.}~\bibnamefont {Ceriotti}},\ }\bibfield  {title} {\bibinfo {title} {Exploring the {Design} {Space} of {Machine} {Learning} {Models} for {Quantum} {Chemistry} with a {Fully} {Differentiable} {Framework}},\ }\bibfield  {journal} {\bibinfo  {journal} {Journal of Chemical Theory and Computation}\ }\href {https://doi.org/10.1021/acs.jctc.5c00522} {10.1021/acs.jctc.5c00522} (\bibinfo {year} {2025}),\ \bibinfo {note} {publisher: American Chemical Society (ACS)}\BibitemShut {NoStop}%
\bibitem [{\citenamefont {Smith}\ \emph {et~al.}(2017{\natexlab{b}})\citenamefont {Smith}, \citenamefont {Isayev},\ and\ \citenamefont {Roitberg}}]{Smith2017b}%
  \BibitemOpen
  \bibfield  {author} {\bibinfo {author} {\bibfnamefont {J.~S.}\ \bibnamefont {Smith}}, \bibinfo {author} {\bibfnamefont {O.}~\bibnamefont {Isayev}},\ and\ \bibinfo {author} {\bibfnamefont {A.~E.}\ \bibnamefont {Roitberg}},\ }\bibfield  {title} {\bibinfo {title} {{ANI}-1: an extensible neural network potential with {DFT} accuracy at force field computational cost},\ }\href {https://doi.org/10.1039/c6sc05720a} {\bibfield  {journal} {\bibinfo  {journal} {Chemical Science}\ }\textbf {\bibinfo {volume} {8}},\ \bibinfo {pages} {3192} (\bibinfo {year} {2017}{\natexlab{b}})},\ \bibinfo {note} {publisher: Royal Society of Chemistry (RSC)}\BibitemShut {NoStop}%
\bibitem [{\citenamefont {Stuke}\ \emph {et~al.}(2019)\citenamefont {Stuke}, \citenamefont {Todorović}, \citenamefont {Rupp}, \citenamefont {Kunkel}, \citenamefont {Ghosh}, \citenamefont {Himanen},\ and\ \citenamefont {Rinke}}]{Stuke2019}%
  \BibitemOpen
  \bibfield  {author} {\bibinfo {author} {\bibfnamefont {A.}~\bibnamefont {Stuke}}, \bibinfo {author} {\bibfnamefont {M.}~\bibnamefont {Todorović}}, \bibinfo {author} {\bibfnamefont {M.}~\bibnamefont {Rupp}}, \bibinfo {author} {\bibfnamefont {C.}~\bibnamefont {Kunkel}}, \bibinfo {author} {\bibfnamefont {K.}~\bibnamefont {Ghosh}}, \bibinfo {author} {\bibfnamefont {L.}~\bibnamefont {Himanen}},\ and\ \bibinfo {author} {\bibfnamefont {P.}~\bibnamefont {Rinke}},\ }\bibfield  {title} {\bibinfo {title} {Chemical diversity in molecular orbital energy predictions with kernel ridge regression},\ }\bibfield  {journal} {\bibinfo  {journal} {The Journal of Chemical Physics}\ }\textbf {\bibinfo {volume} {150}},\ \href {https://doi.org/10.1063/1.5086105} {10.1063/1.5086105} (\bibinfo {year} {2019}),\ \bibinfo {note} {publisher: AIP Publishing}\BibitemShut {NoStop}%
\bibitem [{\citenamefont {Serrano-Morr\'as}\ \emph {et~al.}(2025)\citenamefont {Serrano-Morr\'as}, \citenamefont {Bertran-Mostazo}, \citenamefont {Mi\~narro Lleonar}, \citenamefont {Comajuncosa-Creus}, \citenamefont {Cabello}, \citenamefont {Labranya}, \citenamefont {Escudero}, \citenamefont {Tian}, \citenamefont {Khutorianska}, \citenamefont {Radchenko}, \citenamefont {Moroz}, \citenamefont {Defelipe}, \citenamefont {Ruiz-Carrillo}, \citenamefont {Garcia-Alai}, \citenamefont {Schmidt}, \citenamefont {Rarey}, \citenamefont {Aloy}, \citenamefont {Galdeano}, \citenamefont {Ju\'arez-Jim\'enez},\ and\ \citenamefont {Barril}}]{Serrano-Morras2025}%
  \BibitemOpen
  \bibfield  {author} {\bibinfo {author} {\bibfnamefont {A.}~\bibnamefont {Serrano-Morr\'as}}, \bibinfo {author} {\bibfnamefont {A.}~\bibnamefont {Bertran-Mostazo}}, \bibinfo {author} {\bibfnamefont {M.}~\bibnamefont {Mi\~narro Lleonar}}, \bibinfo {author} {\bibfnamefont {A.}~\bibnamefont {Comajuncosa-Creus}}, \bibinfo {author} {\bibfnamefont {A.}~\bibnamefont {Cabello}}, \bibinfo {author} {\bibfnamefont {C.}~\bibnamefont {Labranya}}, \bibinfo {author} {\bibfnamefont {C.}~\bibnamefont {Escudero}}, \bibinfo {author} {\bibfnamefont {T.~V.}\ \bibnamefont {Tian}}, \bibinfo {author} {\bibfnamefont {I.}~\bibnamefont {Khutorianska}}, \bibinfo {author} {\bibfnamefont {D.~S.}\ \bibnamefont {Radchenko}}, \bibinfo {author} {\bibfnamefont {Y.~S.}\ \bibnamefont {Moroz}}, \bibinfo {author} {\bibfnamefont {L.}~\bibnamefont {Defelipe}}, \bibinfo {author} {\bibfnamefont {D.}~\bibnamefont {Ruiz-Carrillo}}, \bibinfo {author} {\bibfnamefont {M.}~\bibnamefont {Garcia-Alai}}, \bibinfo {author} {\bibfnamefont {R.}~\bibnamefont
  {Schmidt}}, \bibinfo {author} {\bibfnamefont {M.}~\bibnamefont {Rarey}}, \bibinfo {author} {\bibfnamefont {P.}~\bibnamefont {Aloy}}, \bibinfo {author} {\bibfnamefont {C.}~\bibnamefont {Galdeano}}, \bibinfo {author} {\bibfnamefont {J.}~\bibnamefont {Ju\'arez-Jim\'enez}},\ and\ \bibinfo {author} {\bibfnamefont {X.}~\bibnamefont {Barril}},\ }\bibfield  {title} {\bibinfo {title} {A bottom-up approach to find lead compounds in expansive chemical spaces},\ }\href {https://doi.org/10.1038/s42004-025-01610-2} {\bibfield  {journal} {\bibinfo  {journal} {Communications Chemistry}\ }\textbf {\bibinfo {volume} {8}},\ \bibinfo {pages} {225} (\bibinfo {year} {2025})}\BibitemShut {NoStop}%
\bibitem [{\citenamefont {Vogt}(2023)}]{Vogt2023}%
  \BibitemOpen
  \bibfield  {author} {\bibinfo {author} {\bibfnamefont {M.}~\bibnamefont {Vogt}},\ }\bibfield  {title} {\bibinfo {title} {Exploring chemical space — {Generative} models and their evaluation},\ }\href {https://doi.org/10.1016/j.ailsci.2023.100064} {\bibfield  {journal} {\bibinfo  {journal} {Artificial Intelligence in the Life Sciences}\ }\textbf {\bibinfo {volume} {3}},\ \bibinfo {pages} {100064} (\bibinfo {year} {2023})}\BibitemShut {NoStop}%
\bibitem [{\citenamefont {Anstine}\ and\ \citenamefont {Isayev}(2023)}]{Anstine2023}%
  \BibitemOpen
  \bibfield  {author} {\bibinfo {author} {\bibfnamefont {D.~M.}\ \bibnamefont {Anstine}}\ and\ \bibinfo {author} {\bibfnamefont {O.}~\bibnamefont {Isayev}},\ }\bibfield  {title} {\bibinfo {title} {Generative {Models} as an {Emerging} {Paradigm} in the {Chemical} {Sciences}},\ }\href {https://doi.org/10.1021/jacs.2c13467} {\bibfield  {journal} {\bibinfo  {journal} {Journal of the American Chemical Society}\ }\textbf {\bibinfo {volume} {145}},\ \bibinfo {pages} {8736} (\bibinfo {year} {2023})},\ \bibinfo {note} {publisher: American Chemical Society (ACS)}\BibitemShut {NoStop}%
\bibitem [{\citenamefont {Langevin}\ \emph {et~al.}(2023)\citenamefont {Langevin}, \citenamefont {Grebner}, \citenamefont {Güssregen}, \citenamefont {Sauer}, \citenamefont {Li}, \citenamefont {Matter},\ and\ \citenamefont {Bianciotto}}]{Langevin2023}%
  \BibitemOpen
  \bibfield  {author} {\bibinfo {author} {\bibfnamefont {M.}~\bibnamefont {Langevin}}, \bibinfo {author} {\bibfnamefont {C.}~\bibnamefont {Grebner}}, \bibinfo {author} {\bibfnamefont {S.}~\bibnamefont {Güssregen}}, \bibinfo {author} {\bibfnamefont {S.}~\bibnamefont {Sauer}}, \bibinfo {author} {\bibfnamefont {Y.}~\bibnamefont {Li}}, \bibinfo {author} {\bibfnamefont {H.}~\bibnamefont {Matter}},\ and\ \bibinfo {author} {\bibfnamefont {M.}~\bibnamefont {Bianciotto}},\ }\bibfield  {title} {\bibinfo {title} {Impact of {Applicability} {Domains} to {Generative} {Artificial} {Intelligence}},\ }\href {https://doi.org/10.1021/acsomega.3c00883} {\bibfield  {journal} {\bibinfo  {journal} {ACS Omega}\ }\textbf {\bibinfo {volume} {8}},\ \bibinfo {pages} {23148} (\bibinfo {year} {2023})},\ \bibinfo {note} {publisher: American Chemical Society (ACS)}\BibitemShut {NoStop}%
\bibitem [{\citenamefont {Sosnin}(2025)}]{Sosnin2025}%
  \BibitemOpen
  \bibfield  {author} {\bibinfo {author} {\bibfnamefont {S.}~\bibnamefont {Sosnin}},\ }\bibfield  {title} {\bibinfo {title} {Chemical space visual navigation in the era of deep learning and {Big} {Data}},\ }\href {https://doi.org/10.1016/j.drudis.2025.104392} {\bibfield  {journal} {\bibinfo  {journal} {Drug Discovery Today}\ }\textbf {\bibinfo {volume} {30}},\ \bibinfo {pages} {104392} (\bibinfo {year} {2025})}\BibitemShut {NoStop}%
\bibitem [{\citenamefont {Duan}\ \emph {et~al.}(2019{\natexlab{b}})\citenamefont {Duan}, \citenamefont {Janet}, \citenamefont {Liu}, \citenamefont {Nandy},\ and\ \citenamefont {Kulik}}]{Duan2019a}%
  \BibitemOpen
  \bibfield  {author} {\bibinfo {author} {\bibfnamefont {C.}~\bibnamefont {Duan}}, \bibinfo {author} {\bibfnamefont {J.~P.}\ \bibnamefont {Janet}}, \bibinfo {author} {\bibfnamefont {F.}~\bibnamefont {Liu}}, \bibinfo {author} {\bibfnamefont {A.}~\bibnamefont {Nandy}},\ and\ \bibinfo {author} {\bibfnamefont {H.~J.}\ \bibnamefont {Kulik}},\ }\bibfield  {title} {\bibinfo {title} {Learning from {Failure}: {Predicting} {Electronic} {Structure} {Calculation} {Outcomes} with {Machine} {Learning} {Models}},\ }\href {https://doi.org/10.1021/acs.jctc.9b00057} {\bibfield  {journal} {\bibinfo  {journal} {Journal of Chemical Theory and Computation}\ }\textbf {\bibinfo {volume} {15}},\ \bibinfo {pages} {2331} (\bibinfo {year} {2019}{\natexlab{b}})}\BibitemShut {NoStop}%
\bibitem [{\citenamefont {Bugg}\ \emph {et~al.}(1964)\citenamefont {Bugg}, \citenamefont {Desiderato},\ and\ \citenamefont {Sass}}]{Bugg1964}%
  \BibitemOpen
  \bibfield  {author} {\bibinfo {author} {\bibfnamefont {C.}~\bibnamefont {Bugg}}, \bibinfo {author} {\bibfnamefont {R.}~\bibnamefont {Desiderato}},\ and\ \bibinfo {author} {\bibfnamefont {R.~L.}\ \bibnamefont {Sass}},\ }\bibfield  {title} {\bibinfo {title} {An {X}-{Ray} {Diffraction} {Study} of {Nonplanar} {Carbanion} {Structures}},\ }\href {https://doi.org/10.1021/ja01069a040} {\bibfield  {journal} {\bibinfo  {journal} {Journal of the American Chemical Society}\ }\textbf {\bibinfo {volume} {86}},\ \bibinfo {pages} {3157} (\bibinfo {year} {1964})}\BibitemShut {NoStop}%
\bibitem [{\citenamefont {Malischewski}\ and\ \citenamefont {Seppelt}(2017)}]{Malischewski2017}%
  \BibitemOpen
  \bibfield  {author} {\bibinfo {author} {\bibfnamefont {M.}~\bibnamefont {Malischewski}}\ and\ \bibinfo {author} {\bibfnamefont {K.}~\bibnamefont {Seppelt}},\ }\bibfield  {title} {\bibinfo {title} {Crystal {Structure} {Determination} of the {Pentagonal}‐{Pyramidal} {Hexamethylbenzene} {Dication} {C}$_{\textrm{6}}$ ({CH}$_{\textrm{3}}$ )$_{\textrm{6}}$$^{\textrm{2+}}$},\ }\href {https://doi.org/10.1002/anie.201608795} {\bibfield  {journal} {\bibinfo  {journal} {Angewandte Chemie International Edition}\ }\textbf {\bibinfo {volume} {56}},\ \bibinfo {pages} {368} (\bibinfo {year} {2017})}\BibitemShut {NoStop}%
\bibitem [{\citenamefont {Khriachtchev}\ \emph {et~al.}(2000)\citenamefont {Khriachtchev}, \citenamefont {Pettersson}, \citenamefont {Runeberg}, \citenamefont {Lundell},\ and\ \citenamefont {Räsänen}}]{Khriachtchev2000}%
  \BibitemOpen
  \bibfield  {author} {\bibinfo {author} {\bibfnamefont {L.}~\bibnamefont {Khriachtchev}}, \bibinfo {author} {\bibfnamefont {M.}~\bibnamefont {Pettersson}}, \bibinfo {author} {\bibfnamefont {N.}~\bibnamefont {Runeberg}}, \bibinfo {author} {\bibfnamefont {J.}~\bibnamefont {Lundell}},\ and\ \bibinfo {author} {\bibfnamefont {M.}~\bibnamefont {Räsänen}},\ }\bibfield  {title} {\bibinfo {title} {A stable argon compound},\ }\href {https://doi.org/10.1038/35022551} {\bibfield  {journal} {\bibinfo  {journal} {Nature}\ }\textbf {\bibinfo {volume} {406}},\ \bibinfo {pages} {874} (\bibinfo {year} {2000})}\BibitemShut {NoStop}%
\bibitem [{\citenamefont {Qian}\ \emph {et~al.}(2025)\citenamefont {Qian}, \citenamefont {Mardyukov},\ and\ \citenamefont {Schreiner}}]{Qian2025}%
  \BibitemOpen
  \bibfield  {author} {\bibinfo {author} {\bibfnamefont {W.}~\bibnamefont {Qian}}, \bibinfo {author} {\bibfnamefont {A.}~\bibnamefont {Mardyukov}},\ and\ \bibinfo {author} {\bibfnamefont {P.~R.}\ \bibnamefont {Schreiner}},\ }\bibfield  {title} {\bibinfo {title} {Preparation of a neutral nitrogen allotrope hexanitrogen {C2h}-{N6}},\ }\href {https://doi.org/10.1038/s41586-025-09032-9} {\bibfield  {journal} {\bibinfo  {journal} {Nature}\ }\textbf {\bibinfo {volume} {642}},\ \bibinfo {pages} {356} (\bibinfo {year} {2025})}\BibitemShut {NoStop}%
\bibitem [{\citenamefont {Gao}\ \emph {et~al.}(2025)\citenamefont {Gao}, \citenamefont {Gupta}, \citenamefont {Rončević}, \citenamefont {Mycroft}, \citenamefont {Gates}, \citenamefont {Parker},\ and\ \citenamefont {Anderson}}]{Gao2025}%
  \BibitemOpen
  \bibfield  {author} {\bibinfo {author} {\bibfnamefont {Y.}~\bibnamefont {Gao}}, \bibinfo {author} {\bibfnamefont {P.}~\bibnamefont {Gupta}}, \bibinfo {author} {\bibfnamefont {I.}~\bibnamefont {Rončević}}, \bibinfo {author} {\bibfnamefont {C.}~\bibnamefont {Mycroft}}, \bibinfo {author} {\bibfnamefont {P.~J.}\ \bibnamefont {Gates}}, \bibinfo {author} {\bibfnamefont {A.~W.}\ \bibnamefont {Parker}},\ and\ \bibinfo {author} {\bibfnamefont {H.~L.}\ \bibnamefont {Anderson}},\ }\bibfield  {title} {\bibinfo {title} {Solution-phase stabilization of a cyclocarbon by catenane formation},\ }\href {https://doi.org/10.1126/science.ady6054} {\bibfield  {journal} {\bibinfo  {journal} {Science}\ }\textbf {\bibinfo {volume} {389}},\ \bibinfo {pages} {708} (\bibinfo {year} {2025})}\BibitemShut {NoStop}%
\bibitem [{\citenamefont {Benecke}\ \emph {et~al.}(1997)\citenamefont {Benecke}, \citenamefont {Grüner}, \citenamefont {Kerber}, \citenamefont {Laue},\ and\ \citenamefont {Wieland}}]{Benecke1997}%
  \BibitemOpen
  \bibfield  {author} {\bibinfo {author} {\bibfnamefont {C.}~\bibnamefont {Benecke}}, \bibinfo {author} {\bibfnamefont {T.}~\bibnamefont {Grüner}}, \bibinfo {author} {\bibfnamefont {A.}~\bibnamefont {Kerber}}, \bibinfo {author} {\bibfnamefont {R.}~\bibnamefont {Laue}},\ and\ \bibinfo {author} {\bibfnamefont {T.}~\bibnamefont {Wieland}},\ }\bibfield  {title} {\bibinfo {title} {{MOLecular} structure {GENeration} with {MOLGEN}, new features and future developments},\ }\href {https://doi.org/10.1007/s002160050530} {\bibfield  {journal} {\bibinfo  {journal} {Fresenius Journal of Analytical Chemistry}\ }\textbf {\bibinfo {volume} {359}},\ \bibinfo {pages} {23} (\bibinfo {year} {1997})},\ \bibinfo {note} {publisher: Springer Science and Business Media LLC}\BibitemShut {NoStop}%
\bibitem [{\citenamefont {Peironcely}\ \emph {et~al.}(2012)\citenamefont {Peironcely}, \citenamefont {Rojas-Chertó}, \citenamefont {Fichera}, \citenamefont {Reijmers}, \citenamefont {Coulier}, \citenamefont {Faulon},\ and\ \citenamefont {Hankemeier}}]{Peironcely2012}%
  \BibitemOpen
  \bibfield  {author} {\bibinfo {author} {\bibfnamefont {J.~E.}\ \bibnamefont {Peironcely}}, \bibinfo {author} {\bibfnamefont {M.}~\bibnamefont {Rojas-Chertó}}, \bibinfo {author} {\bibfnamefont {D.}~\bibnamefont {Fichera}}, \bibinfo {author} {\bibfnamefont {T.}~\bibnamefont {Reijmers}}, \bibinfo {author} {\bibfnamefont {L.}~\bibnamefont {Coulier}}, \bibinfo {author} {\bibfnamefont {J.-L.}\ \bibnamefont {Faulon}},\ and\ \bibinfo {author} {\bibfnamefont {T.}~\bibnamefont {Hankemeier}},\ }\bibfield  {title} {\bibinfo {title} {{OMG}: {Open} molecule generator},\ }\bibfield  {journal} {\bibinfo  {journal} {Journal of Cheminformatics}\ }\textbf {\bibinfo {volume} {4}},\ \href {https://doi.org/10.1186/1758-2946-4-21} {10.1186/1758-2946-4-21} (\bibinfo {year} {2012}),\ \bibinfo {note} {publisher: Springer Science and Business Media LLC}\BibitemShut {NoStop}%
\bibitem [{\citenamefont {Jaghoori}\ \emph {et~al.}(2013)\citenamefont {Jaghoori}, \citenamefont {Jongmans}, \citenamefont {De~Boer}, \citenamefont {Peironcely}, \citenamefont {Faulon}, \citenamefont {Reijmers},\ and\ \citenamefont {Hankemeier}}]{Jaghoori2013}%
  \BibitemOpen
  \bibfield  {author} {\bibinfo {author} {\bibfnamefont {M.~M.}\ \bibnamefont {Jaghoori}}, \bibinfo {author} {\bibfnamefont {S.-S.~T.}\ \bibnamefont {Jongmans}}, \bibinfo {author} {\bibfnamefont {F.}~\bibnamefont {De~Boer}}, \bibinfo {author} {\bibfnamefont {J.}~\bibnamefont {Peironcely}}, \bibinfo {author} {\bibfnamefont {J.-L.}\ \bibnamefont {Faulon}}, \bibinfo {author} {\bibfnamefont {T.}~\bibnamefont {Reijmers}},\ and\ \bibinfo {author} {\bibfnamefont {T.}~\bibnamefont {Hankemeier}},\ }\bibfield  {title} {\bibinfo {title} {{PMG}: {Multi}-core {Metabolite} {Identification}},\ }\href {https://doi.org/10.1016/j.entcs.2013.11.005} {\bibfield  {journal} {\bibinfo  {journal} {Electronic Notes in Theoretical Computer Science}\ }\textbf {\bibinfo {volume} {299}},\ \bibinfo {pages} {53} (\bibinfo {year} {2013})}\BibitemShut {NoStop}%
\bibitem [{\citenamefont {McKay}\ \emph {et~al.}(2022)\citenamefont {McKay}, \citenamefont {Yirik},\ and\ \citenamefont {Steinbeck}}]{McKay2022}%
  \BibitemOpen
  \bibfield  {author} {\bibinfo {author} {\bibfnamefont {B.~D.}\ \bibnamefont {McKay}}, \bibinfo {author} {\bibfnamefont {M.~A.}\ \bibnamefont {Yirik}},\ and\ \bibinfo {author} {\bibfnamefont {C.}~\bibnamefont {Steinbeck}},\ }\bibfield  {title} {\bibinfo {title} {Surge: a fast open-source chemical graph generator},\ }\bibfield  {journal} {\bibinfo  {journal} {Journal of Cheminformatics}\ }\textbf {\bibinfo {volume} {14}},\ \href {https://doi.org/10.1186/s13321-022-00604-9} {10.1186/s13321-022-00604-9} (\bibinfo {year} {2022}),\ \bibinfo {note} {publisher: Springer Science and Business Media LLC}\BibitemShut {NoStop}%
\bibitem [{\citenamefont {Yirik}\ \emph {et~al.}(2021)\citenamefont {Yirik}, \citenamefont {Sorokina},\ and\ \citenamefont {Steinbeck}}]{Yirik2021}%
  \BibitemOpen
  \bibfield  {author} {\bibinfo {author} {\bibfnamefont {M.~A.}\ \bibnamefont {Yirik}}, \bibinfo {author} {\bibfnamefont {M.}~\bibnamefont {Sorokina}},\ and\ \bibinfo {author} {\bibfnamefont {C.}~\bibnamefont {Steinbeck}},\ }\bibfield  {title} {\bibinfo {title} {{MAYGEN}: an open-source chemical structure generator for constitutional isomers based on the orderly generation principle},\ }\bibfield  {journal} {\bibinfo  {journal} {Journal of Cheminformatics}\ }\textbf {\bibinfo {volume} {13}},\ \href {https://doi.org/10.1186/s13321-021-00529-9} {10.1186/s13321-021-00529-9} (\bibinfo {year} {2021}),\ \bibinfo {note} {publisher: Springer Science and Business Media LLC}\BibitemShut {NoStop}%
\bibitem [{\citenamefont {Rieder}\ \emph {et~al.}(2023)\citenamefont {Rieder}, \citenamefont {Oliveira}, \citenamefont {Riniker},\ and\ \citenamefont {Hünenberger}}]{Rieder2023}%
  \BibitemOpen
  \bibfield  {author} {\bibinfo {author} {\bibfnamefont {S.~R.}\ \bibnamefont {Rieder}}, \bibinfo {author} {\bibfnamefont {M.~P.}\ \bibnamefont {Oliveira}}, \bibinfo {author} {\bibfnamefont {S.}~\bibnamefont {Riniker}},\ and\ \bibinfo {author} {\bibfnamefont {P.~H.}\ \bibnamefont {Hünenberger}},\ }\bibfield  {title} {\bibinfo {title} {Development of an open-source software for isomer enumeration},\ }\href {https://doi.org/10.1186/s13321-022-00677-6} {\bibfield  {journal} {\bibinfo  {journal} {Journal of Cheminformatics}\ }\textbf {\bibinfo {volume} {15}},\ \bibinfo {pages} {10} (\bibinfo {year} {2023})}\BibitemShut {NoStop}%
\bibitem [{\citenamefont {Faulon}(1994)}]{Faulon1994}%
  \BibitemOpen
  \bibfield  {author} {\bibinfo {author} {\bibfnamefont {J.-L.}\ \bibnamefont {Faulon}},\ }\bibfield  {title} {\bibinfo {title} {Stochastic {Generator} of {Chemical} {Structure}. 1. {Application} to the {Structure} {Elucidation} of {Large} {Molecules}},\ }\href {https://doi.org/10.1021/ci00021a031} {\bibfield  {journal} {\bibinfo  {journal} {Journal of Chemical Information and Computer Sciences}\ }\textbf {\bibinfo {volume} {34}},\ \bibinfo {pages} {1204} (\bibinfo {year} {1994})}\BibitemShut {NoStop}%
\bibitem [{\citenamefont {Gasevic}\ \emph {et~al.}(2025)\citenamefont {Gasevic}, \citenamefont {Müller}, \citenamefont {Schöps}, \citenamefont {Lanius}, \citenamefont {Hermann}, \citenamefont {Grimme},\ and\ \citenamefont {Hansen}}]{Gasevic2025}%
  \BibitemOpen
  \bibfield  {author} {\bibinfo {author} {\bibfnamefont {T.}~\bibnamefont {Gasevic}}, \bibinfo {author} {\bibfnamefont {M.}~\bibnamefont {Müller}}, \bibinfo {author} {\bibfnamefont {J.}~\bibnamefont {Schöps}}, \bibinfo {author} {\bibfnamefont {S.}~\bibnamefont {Lanius}}, \bibinfo {author} {\bibfnamefont {J.}~\bibnamefont {Hermann}}, \bibinfo {author} {\bibfnamefont {S.}~\bibnamefont {Grimme}},\ and\ \bibinfo {author} {\bibfnamefont {A.}~\bibnamefont {Hansen}},\ }\href {https://doi.org/10.26434/chemrxiv-2025-rdsd0-v2} {\bibinfo {title} {Chemical {Space} {Exploration} with {Artificial} ”{Mindless}” {Molecules}}} (\bibinfo {year} {2025})\BibitemShut {NoStop}%
\bibitem [{\citenamefont {Karandashev}\ \emph {et~al.}(2023)\citenamefont {Karandashev}, \citenamefont {Weinreich}, \citenamefont {Heinen}, \citenamefont {Arismendi~Arrieta}, \citenamefont {von Rudorff}, \citenamefont {Hermansson},\ and\ \citenamefont {von Lilienfeld}}]{Karandashev2023a}%
  \BibitemOpen
  \bibfield  {author} {\bibinfo {author} {\bibfnamefont {K.}~\bibnamefont {Karandashev}}, \bibinfo {author} {\bibfnamefont {J.}~\bibnamefont {Weinreich}}, \bibinfo {author} {\bibfnamefont {S.}~\bibnamefont {Heinen}}, \bibinfo {author} {\bibfnamefont {D.~J.}\ \bibnamefont {Arismendi~Arrieta}}, \bibinfo {author} {\bibfnamefont {G.~F.}\ \bibnamefont {von Rudorff}}, \bibinfo {author} {\bibfnamefont {K.}~\bibnamefont {Hermansson}},\ and\ \bibinfo {author} {\bibfnamefont {O.~A.}\ \bibnamefont {von Lilienfeld}},\ }\bibfield  {title} {\bibinfo {title} {Evolutionary monte carlo of {QM} properties in chemical space: {Electrolyte} design},\ }\href {https://doi.org/10.1021/acs.jctc.3c00822} {\bibfield  {journal} {\bibinfo  {journal} {Journal of Chemical Theory and Computation}\ }\textbf {\bibinfo {volume} {19}},\ \bibinfo {pages} {8861} (\bibinfo {year} {2023})},\ \bibinfo {note} {publisher: American Chemical Society (ACS)}\BibitemShut {NoStop}%
\bibitem [{\citenamefont {Krenn}\ \emph {et~al.}(2022)\citenamefont {Krenn}, \citenamefont {Ai}, \citenamefont {Barthel}, \citenamefont {Carson}, \citenamefont {Frei}, \citenamefont {Frey}, \citenamefont {Friederich}, \citenamefont {Gaudin}, \citenamefont {Gayle}, \citenamefont {Jablonka}, \citenamefont {Lameiro}, \citenamefont {Lemm}, \citenamefont {Lo}, \citenamefont {Moosavi}, \citenamefont {Nápoles-Duarte}, \citenamefont {Nigam}, \citenamefont {Pollice}, \citenamefont {Rajan}, \citenamefont {Schatzschneider}, \citenamefont {Schwaller}, \citenamefont {Skreta}, \citenamefont {Smit}, \citenamefont {Strieth-Kalthoff}, \citenamefont {Sun}, \citenamefont {Tom}, \citenamefont {von Rudorff}, \citenamefont {Wang}, \citenamefont {White}, \citenamefont {Young}, \citenamefont {Yu},\ and\ \citenamefont {Aspuru-Guzik}}]{Krenn2022}%
  \BibitemOpen
  \bibfield  {author} {\bibinfo {author} {\bibfnamefont {M.}~\bibnamefont {Krenn}}, \bibinfo {author} {\bibfnamefont {Q.}~\bibnamefont {Ai}}, \bibinfo {author} {\bibfnamefont {S.}~\bibnamefont {Barthel}}, \bibinfo {author} {\bibfnamefont {N.}~\bibnamefont {Carson}}, \bibinfo {author} {\bibfnamefont {A.}~\bibnamefont {Frei}}, \bibinfo {author} {\bibfnamefont {N.~C.}\ \bibnamefont {Frey}}, \bibinfo {author} {\bibfnamefont {P.}~\bibnamefont {Friederich}}, \bibinfo {author} {\bibfnamefont {T.}~\bibnamefont {Gaudin}}, \bibinfo {author} {\bibfnamefont {A.~A.}\ \bibnamefont {Gayle}}, \bibinfo {author} {\bibfnamefont {K.~M.}\ \bibnamefont {Jablonka}}, \bibinfo {author} {\bibfnamefont {R.~F.}\ \bibnamefont {Lameiro}}, \bibinfo {author} {\bibfnamefont {D.}~\bibnamefont {Lemm}}, \bibinfo {author} {\bibfnamefont {A.}~\bibnamefont {Lo}}, \bibinfo {author} {\bibfnamefont {S.~M.}\ \bibnamefont {Moosavi}}, \bibinfo {author} {\bibfnamefont {J.~M.}\ \bibnamefont {Nápoles-Duarte}}, \bibinfo {author} {\bibfnamefont
  {A.}~\bibnamefont {Nigam}}, \bibinfo {author} {\bibfnamefont {R.}~\bibnamefont {Pollice}}, \bibinfo {author} {\bibfnamefont {K.}~\bibnamefont {Rajan}}, \bibinfo {author} {\bibfnamefont {U.}~\bibnamefont {Schatzschneider}}, \bibinfo {author} {\bibfnamefont {P.}~\bibnamefont {Schwaller}}, \bibinfo {author} {\bibfnamefont {M.}~\bibnamefont {Skreta}}, \bibinfo {author} {\bibfnamefont {B.}~\bibnamefont {Smit}}, \bibinfo {author} {\bibfnamefont {F.}~\bibnamefont {Strieth-Kalthoff}}, \bibinfo {author} {\bibfnamefont {C.}~\bibnamefont {Sun}}, \bibinfo {author} {\bibfnamefont {G.}~\bibnamefont {Tom}}, \bibinfo {author} {\bibfnamefont {G.~F.}\ \bibnamefont {von Rudorff}}, \bibinfo {author} {\bibfnamefont {A.}~\bibnamefont {Wang}}, \bibinfo {author} {\bibfnamefont {A.~D.}\ \bibnamefont {White}}, \bibinfo {author} {\bibfnamefont {A.}~\bibnamefont {Young}}, \bibinfo {author} {\bibfnamefont {R.}~\bibnamefont {Yu}},\ and\ \bibinfo {author} {\bibfnamefont {A.}~\bibnamefont {Aspuru-Guzik}},\ }\bibfield  {title} {\bibinfo
  {title} {{SELFIES} and the future of molecular string representations},\ }\href {https://doi.org/10.1016/j.patter.2022.100588} {\bibfield  {journal} {\bibinfo  {journal} {Patterns}\ }\textbf {\bibinfo {volume} {3}},\ \bibinfo {pages} {100588} (\bibinfo {year} {2022})},\ \bibinfo {note} {publisher: Elsevier BV}\BibitemShut {NoStop}%
\bibitem [{\citenamefont {Watts}\ and\ \citenamefont {Strogatz}(1998)}]{Watts1998}%
  \BibitemOpen
  \bibfield  {author} {\bibinfo {author} {\bibfnamefont {D.~J.}\ \bibnamefont {Watts}}\ and\ \bibinfo {author} {\bibfnamefont {S.~H.}\ \bibnamefont {Strogatz}},\ }\bibfield  {title} {\bibinfo {title} {Collective dynamics of ‘small-world’ networks},\ }\href {https://doi.org/10.1038/30918} {\bibfield  {journal} {\bibinfo  {journal} {Nature}\ }\textbf {\bibinfo {volume} {393}},\ \bibinfo {pages} {440} (\bibinfo {year} {1998})}\BibitemShut {NoStop}%
\bibitem [{\citenamefont {Greenhill}\ and\ \citenamefont {McKay}(2013)}]{Greenhill2013}%
  \BibitemOpen
  \bibfield  {author} {\bibinfo {author} {\bibfnamefont {C.}~\bibnamefont {Greenhill}}\ and\ \bibinfo {author} {\bibfnamefont {B.~D.}\ \bibnamefont {McKay}},\ }\bibfield  {title} {\bibinfo {title} {Asymptotic enumeration of sparse multigraphs with given degrees},\ }\href {https://doi.org/10.1137/130913419} {\bibfield  {journal} {\bibinfo  {journal} {SIAM Journal on Discrete Mathematics}\ }\textbf {\bibinfo {volume} {27}},\ \bibinfo {pages} {2064} (\bibinfo {year} {2013})},\ \bibinfo {note} {publisher: Society for Industrial \& Applied Mathematics (SIAM)}\BibitemShut {NoStop}%
\bibitem [{\citenamefont {Croes}(1958)}]{Croes1958}%
  \BibitemOpen
  \bibfield  {author} {\bibinfo {author} {\bibfnamefont {G.~A.}\ \bibnamefont {Croes}},\ }\bibfield  {title} {\bibinfo {title} {A {Method} for {Solving} {Traveling}-{Salesman} {Problems}},\ }\href {http://www.jstor.org/stable/167074} {\bibfield  {journal} {\bibinfo  {journal} {Operations Research}\ }\textbf {\bibinfo {volume} {6}},\ \bibinfo {pages} {791} (\bibinfo {year} {1958})},\ \bibinfo {note} {publisher: INFORMS}\BibitemShut {NoStop}%
\bibitem [{\citenamefont {Abu-Aisheh}\ \emph {et~al.}(2015)\citenamefont {Abu-Aisheh}, \citenamefont {Raveaux}, \citenamefont {Ramel},\ and\ \citenamefont {Martineau}}]{Abu-Aisheh2015}%
  \BibitemOpen
  \bibfield  {author} {\bibinfo {author} {\bibfnamefont {Z.}~\bibnamefont {Abu-Aisheh}}, \bibinfo {author} {\bibfnamefont {R.}~\bibnamefont {Raveaux}}, \bibinfo {author} {\bibfnamefont {J.-Y.}\ \bibnamefont {Ramel}},\ and\ \bibinfo {author} {\bibfnamefont {P.}~\bibnamefont {Martineau}},\ }\bibfield  {title} {\bibinfo {title} {An {Exact} {Graph} {Edit} {Distance} {Algorithm} for {Solving} {Pattern} {Recognition} {Problems}:},\ }in\ \href {https://doi.org/10.5220/0005209202710278} {\emph {\bibinfo {booktitle} {Proceedings of the {International} {Conference} on {Pattern} {Recognition} {Applications} and {Methods}}}}\ (\bibinfo  {publisher} {SCITEPRESS - Science and and Technology Publications},\ \bibinfo {address} {Lisbon, Portugal},\ \bibinfo {year} {2015})\ pp.\ \bibinfo {pages} {271--278}\BibitemShut {NoStop}%
\bibitem [{\citenamefont {Banjafar}\ \emph {et~al.}(2025)\citenamefont {Banjafar}, \citenamefont {Monterrubio-Chanca},\ and\ \citenamefont {von Rudorff}}]{Banjafar2025c}%
  \BibitemOpen
  \bibfield  {author} {\bibinfo {author} {\bibfnamefont {A.}~\bibnamefont {Banjafar}}, \bibinfo {author} {\bibfnamefont {D.~J.}\ \bibnamefont {Monterrubio-Chanca}},\ and\ \bibinfo {author} {\bibfnamefont {G.~F.}\ \bibnamefont {von Rudorff}},\ }\href {https://doi.org/10.5281/ZENODO.16980244} {\bibinfo {title} {{NablaChem}/nablachem: v25.8}} (\bibinfo {year} {2025})\BibitemShut {NoStop}%
\bibitem [{\citenamefont {Greenhill}(2015)}]{Greenhill2015}%
  \BibitemOpen
  \bibfield  {author} {\bibinfo {author} {\bibfnamefont {C.}~\bibnamefont {Greenhill}},\ }\bibfield  {title} {\bibinfo {title} {The switch {Markov} chain for sampling irregular graphs ({Extended} {Abstract})},\ }in\ \href {https://doi.org/10.1137/1.9781611973730.103} {\emph {\bibinfo {booktitle} {Proceedings of the {Twenty}-{Sixth} {Annual} {ACM}-{SIAM} {Symposium} on {Discrete} {Algorithms}}}}\ (\bibinfo  {publisher} {Society for Industrial and Applied Mathematics},\ \bibinfo {year} {2015})\ pp.\ \bibinfo {pages} {1564--1572}\BibitemShut {NoStop}%
\bibitem [{\citenamefont {jamesross2}(2023)}]{jamesross22023}%
  \BibitemOpen
  \bibfield  {author} {\bibinfo {author} {\bibnamefont {jamesross2}},\ }\href {https://github.com/jamesross2/random_graph} {\bibinfo {title} {jamesross2/random\_graph}} (\bibinfo {year} {2023}),\ \bibinfo {note} {original-date: 2020-04-05T22:22:04Z}\BibitemShut {NoStop}%
\bibitem [{\citenamefont {Smith}\ \emph {et~al.}(2017{\natexlab{c}})\citenamefont {Smith}, \citenamefont {Isayev},\ and\ \citenamefont {Roitberg}}]{Smith2017}%
  \BibitemOpen
  \bibfield  {author} {\bibinfo {author} {\bibfnamefont {J.~S.}\ \bibnamefont {Smith}}, \bibinfo {author} {\bibfnamefont {O.}~\bibnamefont {Isayev}},\ and\ \bibinfo {author} {\bibfnamefont {A.~E.}\ \bibnamefont {Roitberg}},\ }\bibfield  {title} {\bibinfo {title} {{ANI}-1, {A} data set of 20 million calculated off-equilibrium conformations for organic molecules},\ }\href {https://doi.org/10.1038/sdata.2017.193} {\bibfield  {journal} {\bibinfo  {journal} {Scientific Data}\ }\textbf {\bibinfo {volume} {4}},\ \bibinfo {pages} {170193} (\bibinfo {year} {2017}{\natexlab{c}})}\BibitemShut {NoStop}%
\bibitem [{\citenamefont {Blum}\ and\ \citenamefont {Reymond}(2009{\natexlab{b}})}]{Blum2009c}%
  \BibitemOpen
  \bibfield  {author} {\bibinfo {author} {\bibfnamefont {L.~C.}\ \bibnamefont {Blum}}\ and\ \bibinfo {author} {\bibfnamefont {J.-L.}\ \bibnamefont {Reymond}},\ }\bibfield  {title} {\bibinfo {title} {970 {Million} {Druglike} {Small} {Molecules} for {Virtual} {Screening} in the {Chemical} {Universe} {Database} {GDB}-13},\ }\href {https://doi.org/10.1021/ja902302h} {\bibfield  {journal} {\bibinfo  {journal} {Journal of the American Chemical Society}\ }\textbf {\bibinfo {volume} {131}},\ \bibinfo {pages} {8732} (\bibinfo {year} {2009}{\natexlab{b}})}\BibitemShut {NoStop}%
\bibitem [{\citenamefont {Massey}(1951)}]{Massey1951}%
  \BibitemOpen
  \bibfield  {author} {\bibinfo {author} {\bibfnamefont {F.~J.}\ \bibnamefont {Massey}},\ }\bibfield  {title} {\bibinfo {title} {The {Kolmogorov}-{Smirnov} {Test} for {Goodness} of {Fit}},\ }\href {https://doi.org/10.2307/2280095} {\bibfield  {journal} {\bibinfo  {journal} {Journal of the American Statistical Association}\ }\textbf {\bibinfo {volume} {46}},\ \bibinfo {pages} {68} (\bibinfo {year} {1951})},\ \bibinfo {note} {publisher: JSTOR}\BibitemShut {NoStop}%
\bibitem [{\citenamefont {Kullback}\ and\ \citenamefont {Leibler}(1951)}]{Kullback1951}%
  \BibitemOpen
  \bibfield  {author} {\bibinfo {author} {\bibfnamefont {S.}~\bibnamefont {Kullback}}\ and\ \bibinfo {author} {\bibfnamefont {R.~A.}\ \bibnamefont {Leibler}},\ }\bibfield  {title} {\bibinfo {title} {On {Information} and {Sufficiency}},\ }\href {https://doi.org/10.1214/aoms/1177729694} {\bibfield  {journal} {\bibinfo  {journal} {The Annals of Mathematical Statistics}\ }\textbf {\bibinfo {volume} {22}},\ \bibinfo {pages} {79} (\bibinfo {year} {1951})},\ \bibinfo {note} {publisher: Institute of Mathematical Statistics}\BibitemShut {NoStop}%
\bibitem [{\citenamefont {Rassokhin}\ and\ \citenamefont {Agrafiotis}(2000)}]{Rassokhin2000}%
  \BibitemOpen
  \bibfield  {author} {\bibinfo {author} {\bibfnamefont {D.~N.}\ \bibnamefont {Rassokhin}}\ and\ \bibinfo {author} {\bibfnamefont {D.~K.}\ \bibnamefont {Agrafiotis}},\ }\bibfield  {title} {\bibinfo {title} {Kolmogorov-{Smirnov} statistic and its application in library design},\ }\href {https://doi.org/10.1016/S1093-3263(00)00063-2} {\bibfield  {journal} {\bibinfo  {journal} {Journal of Molecular Graphics and Modelling}\ }\textbf {\bibinfo {volume} {18}},\ \bibinfo {pages} {368} (\bibinfo {year} {2000})}\BibitemShut {NoStop}%
\bibitem [{\citenamefont {Ji}\ \emph {et~al.}(2022)\citenamefont {Ji}, \citenamefont {Zhang}, \citenamefont {Ying}, \citenamefont {Wang}, \citenamefont {Zhao},\ and\ \citenamefont {Gao}}]{Ji2022}%
  \BibitemOpen
  \bibfield  {author} {\bibinfo {author} {\bibfnamefont {S.}~\bibnamefont {Ji}}, \bibinfo {author} {\bibfnamefont {Z.}~\bibnamefont {Zhang}}, \bibinfo {author} {\bibfnamefont {S.}~\bibnamefont {Ying}}, \bibinfo {author} {\bibfnamefont {L.}~\bibnamefont {Wang}}, \bibinfo {author} {\bibfnamefont {X.}~\bibnamefont {Zhao}},\ and\ \bibinfo {author} {\bibfnamefont {Y.}~\bibnamefont {Gao}},\ }\bibfield  {title} {\bibinfo {title} {Kullback–{Leibler} {Divergence} {Metric} {Learning}},\ }\href {https://doi.org/10.1109/TCYB.2020.3008248} {\bibfield  {journal} {\bibinfo  {journal} {IEEE Transactions on Cybernetics}\ }\textbf {\bibinfo {volume} {52}},\ \bibinfo {pages} {2047} (\bibinfo {year} {2022})}\BibitemShut {NoStop}%
\bibitem [{\citenamefont {McClendon}\ \emph {et~al.}(2012)\citenamefont {McClendon}, \citenamefont {Hua}, \citenamefont {Barreiro},\ and\ \citenamefont {Jacobson}}]{McClendon2012a}%
  \BibitemOpen
  \bibfield  {author} {\bibinfo {author} {\bibfnamefont {C.~L.}\ \bibnamefont {McClendon}}, \bibinfo {author} {\bibfnamefont {L.}~\bibnamefont {Hua}}, \bibinfo {author} {\bibfnamefont {G.}~\bibnamefont {Barreiro}},\ and\ \bibinfo {author} {\bibfnamefont {M.~P.}\ \bibnamefont {Jacobson}},\ }\bibfield  {title} {\bibinfo {title} {Comparing {Conformational} {Ensembles} {Using} the {Kullback}–{Leibler} {Divergence} {Expansion}},\ }\href {https://doi.org/10.1021/ct300008d} {\bibfield  {journal} {\bibinfo  {journal} {Journal of Chemical Theory and Computation}\ }\textbf {\bibinfo {volume} {8}},\ \bibinfo {pages} {2115} (\bibinfo {year} {2012})}\BibitemShut {NoStop}%
\bibitem [{\citenamefont {Lagarde}\ \emph {et~al.}(2015)\citenamefont {Lagarde}, \citenamefont {Zagury},\ and\ \citenamefont {Montes}}]{Lagarde2015}%
  \BibitemOpen
  \bibfield  {author} {\bibinfo {author} {\bibfnamefont {N.}~\bibnamefont {Lagarde}}, \bibinfo {author} {\bibfnamefont {J.-F.}\ \bibnamefont {Zagury}},\ and\ \bibinfo {author} {\bibfnamefont {M.}~\bibnamefont {Montes}},\ }\bibfield  {title} {\bibinfo {title} {Benchmarking {Data} {Sets} for the {Evaluation} of {Virtual} {Ligand} {Screening} {Methods}: {Review} and {Perspectives}},\ }\href {https://doi.org/10.1021/acs.jcim.5b00090} {\bibfield  {journal} {\bibinfo  {journal} {Journal of Chemical Information and Modeling}\ }\textbf {\bibinfo {volume} {55}},\ \bibinfo {pages} {1297} (\bibinfo {year} {2015})},\ \bibinfo {note} {publisher: American Chemical Society (ACS)}\BibitemShut {NoStop}%
\bibitem [{\citenamefont {Antoniuk}\ \emph {et~al.}(2025)\citenamefont {Antoniuk}, \citenamefont {Zaman}, \citenamefont {Ben-Nun}, \citenamefont {Li}, \citenamefont {Diffenderfer}, \citenamefont {Demirci}, \citenamefont {Smolenski}, \citenamefont {Hsu}, \citenamefont {Hiszpanski}, \citenamefont {Chiu}, \citenamefont {Kailkhura},\ and\ \citenamefont {Van~Essen}}]{Antoniuk2025}%
  \BibitemOpen
  \bibfield  {author} {\bibinfo {author} {\bibfnamefont {E.~R.}\ \bibnamefont {Antoniuk}}, \bibinfo {author} {\bibfnamefont {S.}~\bibnamefont {Zaman}}, \bibinfo {author} {\bibfnamefont {T.}~\bibnamefont {Ben-Nun}}, \bibinfo {author} {\bibfnamefont {P.}~\bibnamefont {Li}}, \bibinfo {author} {\bibfnamefont {J.}~\bibnamefont {Diffenderfer}}, \bibinfo {author} {\bibfnamefont {B.}~\bibnamefont {Demirci}}, \bibinfo {author} {\bibfnamefont {O.}~\bibnamefont {Smolenski}}, \bibinfo {author} {\bibfnamefont {T.}~\bibnamefont {Hsu}}, \bibinfo {author} {\bibfnamefont {A.~M.}\ \bibnamefont {Hiszpanski}}, \bibinfo {author} {\bibfnamefont {K.}~\bibnamefont {Chiu}}, \bibinfo {author} {\bibfnamefont {B.}~\bibnamefont {Kailkhura}},\ and\ \bibinfo {author} {\bibfnamefont {B.}~\bibnamefont {Van~Essen}},\ }\href {https://doi.org/10.48550/ARXIV.2505.01912} {\bibinfo {title} {{BOOM}: {Benchmarking} {Out}-{Of}-distribution {Molecular} {Property} {Predictions} of {Machine} {Learning} {Models}}} (\bibinfo {year} {2025}),\ \bibinfo
  {note} {version Number: 1}\BibitemShut {NoStop}%
\end{thebibliography}%
\end{document}